\documentclass[12pt]{article}

\usepackage{amsthm}
\usepackage{natbib}
\usepackage{color}
\usepackage{graphicx}
\usepackage{amsmath}
\usepackage{amsfonts}
\usepackage{amssymb}
\usepackage{amsxtra}
\usepackage{amstext}
\usepackage{latexsym}
\usepackage{dsfont}
\usepackage{stmaryrd}
\usepackage{mathrsfs}
\usepackage{bbm,bm}
\usepackage{subfigure}
\usepackage{here}
\usepackage{multirow}
\usepackage{booktabs}
\usepackage{epsfig}
\usepackage{lineno}
\usepackage{hyperref}
\usepackage{url}
\usepackage{soul} %Para tachar usar \textst{}

\theoremstyle{plain}
\newtheorem{theorem}{Theorem}[section]

\theoremstyle{definition}
\newtheorem{remark}{Remark}[section]

\begin{document}
%%==============================================================================================================%%
%%==============================================================================================================%%

\def\spacingset#1{\renewcommand{\baselinestretch}%
{#1}\small\normalsize} \spacingset{1}

%%===========================================%%

\noindent {\textbf{\Large New Mixed Portmanteau Tests for Time Series Models}}

%\title{\textbf{On Mixed Portmanteau Tests for Time Series Models}}
%\maketitle

\vskip 5mm
\noindent Esam Mahdi

\noindent School of Mathematics and Statistics, Carleton University, Ottawa, ON, Canada

\noindent esammahdi@cunet.carleton.ca

%\noindent Corresponding author

\vskip 0.1mm
\centerline{\rule{17cm}{0.4pt}}
\vskip 0.1mm

\section*{\centering \small {Highlights}}

\begin{itemize}
\item
We propose omnibus portmanteau statistics that can be used as a goodness-of-fit and test for the 
presence of (possibly nonlinear) dependence structure in the residuals of time series models.

\item 
The test statistics are based on combining the autocorrelations of the residuals at different powers.

\item 
The asymptotic distribution is derived under the general class of time series models, including ARMA and GARCH, and other nonlinear structures.

\item 
Extensive simulations are conducted to demonstrate the good performance of the proposed tests. 

\item
The proposed tests are implemented in \texttt{R}.

\item
Real time series data are employed to show the practical use of the proposed tests.

\end{itemize}
\vskip 0.1mm
\centerline{\rule{17cm}{0.4pt}}
\vskip 0.1mm

\spacingset{1.3} % DON'T change the spacing!
\bigskip
\begin{abstract}
This article proposes omnibus portmanteau tests for contrasting adequacy of time series models.
The test statistics are based on combining the autocorrelation function of the conditional residuals, the autocorrelation function of the conditional squared residuals, and the cross-correlation function between these residuals and their squares. 
The maximum likelihood estimator is used to derive the asymptotic distribution of the proposed test statistics under a general class of time series models, including ARMA, GARCH, and other nonlinear structures.
An extensive Monte Carlo simulation study shows that the proposed tests successfully control the type I error probability and tend to have more power than other competitor tests in many scenarios. 
Two applications to a set of weekly stock returns for $92$ companies from the S\&P 500 demonstrate the practical use of the proposed tests.
\end{abstract}

\noindent%
{\it Keywords:} ARMA and GARCH models; Autocorrelation; Cross-correlation; Nonlinearity test; Mixed portmanteau tests;
Quasi-maximum likelihood estimation; Stock returns.
\vfill

\newpage
\spacingset{1.3} % DON'T change the spacing!

%%==============================================================================================================%%
\section{\label{sec:introduction}Introduction}
%%==============================================================================================================%%
%\linenumbers

Time series models often consist of two components: (i) the conditional mean part; and (ii) the conditional variance part.
Traditionally, the autoregressive and moving average (ARMA) models are classified as linear and specify the mean part, whereas the generalized autoregressive conditional heteroscedasticity (GARCH) models are nonlinear and describe the variance part.
During the past four decades, time series analysis was dominated by the ARMA models, where a good model should be able to specify the dependence structure of the series adequately \citep{BoxJenkins1970}.
Dependency, in ARMA models, is often measured by using the residual autocorrelation function (ACF). 
To test the adequacy of an ARMA model, a portmanteau statistic was proposed by \cite{BoxPierce1970} based on the distribution of the residual ACF.
Since then, several authors have improved the portmanteau tests \citep[see, for example,][]{FisherGallagher2012,LjungBox1978, PR2002, PR2006, Mahdi2017}.

In the last two decades, the analysis of nonlinear time series models has attracted a great deal of interest in business, economics, finance, and other fields. 
\cite{BoxJenkins1970}, \cite{Granger1978} and \cite{TongLim1980} noticed that the squared residuals of time series models are significantly autocorrelated even though the residuals are not autocorrelated.
This indicates that the error term of these models might be uncorrelated but not independent. 
The authors suggested using the ACF of the squared values of the series to detect nonlinearity.  
In this respect, \cite{Engle1982} showed that the classical portmanteau tests proposed by \cite{BoxPierce1970} and \cite{LjungBox1978} fail to detect the presence of the autoregressive conditional heteroscedasticity (ARCH) in many financial time series models. 
To test for the presence of an ARCH process, \cite{Engle1982} introduced a Lagrange multiplier statistic based on the autocorrelations of the squared residuals. 

Several authors have developed portmanteau test statistics employing the ACF of the squared residuals to detect nonlinear structures and ARCH effect in time series models \citep[see, for example,][]{FisherGallagher2012,McLeodLi1983,PR2002,PR2006,PR2005}.
All the above test statistics were derived under the assumptions of ARMA models and were not proposed for nonlinear time series models.

A portmanteau test was developed by \cite{LiMak1994} to check the adequacy of nonlinear 
time series models, including ARMA-ARCH, and other conditional heteroscedastic structures.
Under a general class of time series models, 
two mixed portmanteau tests, to detect the linear and nonlinear dependency in time series models, were considered by \cite{Wong2005}
 summing the statistics derived by \cite{BoxPierce1970} and \cite{LiMak1994} for the first one, and summing the statistics proposed by \cite{LjungBox1978} and \cite{McLeodLi1983} for the second one.
\cite{Wong2005} showed that their mixed tests are, in many situations, more powerful than the tests proposed by \cite{LjungBox1978} and \cite{McLeodLi1983}, when the fitted model has a disparity in its first and second moments. 
\citet{ZhuKe2013} proposed another mixed portmanteau test for ARMA-GARCH models with parameters estimated by a quasi-maximum exponential likelihood estimator.
\cite{LI20181} proposed a first-order zero-drift GARCH (ZD-GARCH(1, 1)) model to study conditional
heteroscedasticity and heteroscedasticity together, for which the authors constructed a portmanteau test for model checking. Their test statistic was derived based on the lag-$k$ autocorrelation function of the $s$th power of the absolute residuals, where $k$ is a positive integer and $s>0$.

For the test statistics presented by \cite{Wong2005,ZhuKe2013} and \citet{LI20181}, the authors did not consider the cross-correlation between the residuals at different powers. 
The idea of using the cross-correlation between the residuals at different powers to test for linearity was considered by \citet{WelshJernigan1983,LL1985,LL1987,ZM2019}.

In this article, we propose four mixed portmanteau statistics for time series models.
The proposed test statistics are composed by three components: The first of them utilizes the autocorrelations of the residuals, which is  designed to capture the linear dependency in the mean part of time series models. 
Then, the second component of these statistics utilizes the autocorrelation of the squared residuals, which can be used to test for conditional
heteroscedastic effects. 
The third component of these statistics is related to the cross-correlations between the residuals and their squared values, which may be helpful to test for other types of nonlinear models in which the residuals and their squared values are cross-correlated.
The cross-correlations between the residuals and their squared values allow us to propose two different tests. 
One of these tests is based on the positive lags and the other one uses the negative lags. 
Therefore, the tests  proposed in the present study combine the statistics presented in \cite{Wong2005} and \cite{ZM2019}.  

The remainder of this article is organized as follows.
Section \ref{sec:model.Assumptions} defines some popular time series models with their assumptions. 
In Section \ref{sec:new.test}, we propose new auto-and-cross-correlated test statistics for contrasting the adequacy of fitted time series models and derive their asymptotic distributions.
In Section \ref{sec:simulation}, a Monte Carlo simulation study is conducted to compare the performance of the proposed statistics with some tests commonly used in the literature.
We show that the empirical size of the proposed tests successfully controls the type I error probability and tends to have higher power than other tests in many cases.
Section \ref{sec:application} presents illustrative applications to demonstrate the usefulness of the proposed tests for real-world data. 
We finish this article in Section \ref{sec:discussion} by providing concluding remarks.

%%==============================================================================================================%%
\section{\label{sec:model.Assumptions}The general time series model and its assumptions}
%%==============================================================================================================%%

Assume that $\{z_t\mbox{: } t=0,\pm 1,\cdots\}$ is a time series that is generated by the strictly stationary and ergodic model defined by 
\begin{equation}\label{AR.ARCH.model}
	z_t=\mu_{t}(\bm{\theta},\mathcal{F}_{t-1}) +\varepsilon_{t}, \qquad \varepsilon_{t}=\xi_t\sqrt{h_{t}(\bm{\theta})}
\end{equation}
where $\mathcal{F}_{t-1}$ represents the information set ($\sigma$-algebra) generated by $\{z_t,z_{t-1},\cdots\}$, and
$\bm{\theta}$ denotes the $l\times 1$ vector of unknown parameters and its true value is $\bm{\theta}_0$.
$\mu_{t}(\bm{\theta})=\mu_{t}(\bm{\theta},\mathcal{F}_{t-1})=\mathrm{E}(z_t|\mathcal{F}_{t-1})$ and
$h_{t}(\bm{\theta})=\mathrm{Var}(\varepsilon_t|\mathcal{F}_{t-1})>0$ are the conditional mean and conditional variance of $z_t$, respectively.
Both are assumed to have continuous second order derivative almost surely (a.s.). 
The process $\{\xi_t\}$ is a sequence of independent and identically distributed (i.i.d.) random variables with mean zero, variance one, and finite fourth moment.

The usual ARMA-GARCH model can be seen as a special case of this model that can be written as
\begin{align}\label{ARMAGARCH.model}
\nonumber z_t &= \sum_{i =1}^{p}\phi_i z_{t-i} + \sum_{i =0}^{q}\theta_i \varepsilon_{t-i} + \varepsilon_{t}\\
	\varepsilon_t & =\xi_{t}\sqrt{h_{t}(\bm{\theta})},\quad h_t(\bm{\theta})=\omega+\sum_{i=1}^{a}\alpha_i\varepsilon_{t-i}^2(\bm{\theta})+\sum_{j=1}^{b}\beta_j h_{t-j}(\bm{\theta}),
\end{align}
where $\{\xi_t\}$ is a sequence of i.i.d.\,random variables with mean zero, variance one, and {$\mathrm{E}(\xi_t^4)<\infty$, with $\omega>0$,
	$\alpha_i\ge 0$, $\beta_j\ge 0$, for $i\in\{1,\cdots,a\}, j\in\{1,\cdots,b\}$}, and $\sum_{i=1}^{a}\alpha_i+\sum_{j=1}^{b}\beta_j<1$. 

Ignoring the constant term, the Gaussian log-likelihood function of $\{z_1,\cdots,z_n\}$ given the initial values $\{z_t \mbox{: }  t\in\mathbb{Z}^{-}\cup\{0\}\}$ can be written as
{
\begin{equation}\label{log.lik}
\ell(\bm{\theta})=\sum_{t=1}^{n}\ell_{t}(\bm{\theta},z^\star),
\end{equation}
where 
$$
\ell_{t}(\bm{\theta},z^\star)=-\frac{1}{2}\log{(h_t(\bm{\theta}))}-\frac{\varepsilon_{t}^2(\bm{\theta})}{2h_t(\bm{\theta})},\quad t \in \{1,\cdots,n\},
$$
where $z^\star\equiv\{z_t,z_{t-1},\cdots\}$}. 
Assuming the parameter space is $\Theta$, where $\bm{\theta}_0$ is an interior vector in $\Theta$,
and and for convenience, let's denote $\varepsilon_{t}=\varepsilon_{t}(\bm{\theta}_0)$, $\mu_{t}=\mu_{t}(\bm{\theta}_0)$, and $h_t=h_t(\bm{\theta}_0)$.  
The first derivative of the log-likelihood function is given by
$$
\frac{\partial \ell(\bm{\theta}_0)}{\partial\bm{\theta}}=\frac{1}{2}\sum_{t=1}^{n}\frac{1}{h_{t}}\frac{\partial h_{t}}{\partial\bm{\theta}}\left(\frac{\varepsilon_{t}^2}{h_{t}}-1\right)+\sum_{t=1}^{n}\frac{\varepsilon_{t}}{h_{t}}\frac{\partial\mu_{t}}{\partial\bm{\theta}}.
$$
By taking the conditional expectations of the iterative second derivatives with respect to $\mathcal{F}_{t-1}$, we have  
$$\mathrm{E}\left[
\frac{\partial^2 \ell(\bm{\theta}_0)}{\partial\bm{\theta}\partial\bm{\theta}^\top}\right]=-\frac{1}{2}\sum_{t=1}^{n}\frac{1}{h_{t}^2}\mathrm{E}\left(\frac{1}{h_{t}^2}\left(\frac{\partial h_{t}}{\partial\bm{\theta}}\right)\left(\frac{\partial h_{t}}{\partial\bm{\theta}}\right)^\top\right)-
\sum_{t=1}^{n}\frac{1}{h_{t}} \mathrm{E}\left(\frac{1}{h_{t}}\left(\frac{\partial \mu_{t}}{\partial\bm{\theta}}\right)\left(\frac{\partial \mu_{t}}{\partial\bm{\theta}}\right)^\top\right).
$$
Assume that 
$\partial \ell(\bm{\theta}_0)/\partial\bm{\theta}$ is a martingale difference in terms of $\mathcal{F}_{t-1}$ 
and let $\widehat{\bm{\theta}}_n$ be the quasi-maximum likelihood estimator of $\bm{\theta}_0$, 
that is,
$\widehat{\bm{\theta}}_n\overset{\mathrm{a.s.}}{\to}{\bm{\theta}_0}$, where $\overset{\mathrm{a.s.}}{\to}$ denotes a.s.\,convergence.
Then, it follows that 
\begin{equation}\label{eq:quasi}
\sqrt{n}(\widehat{\bm{\theta}}_n-\bm{\theta}_0)=-\frac{1}{\sqrt{n}} \bm \Sigma^{-1}\frac{\partial\ell(\bm{\theta}_0)}{\partial\bm{\theta}} + o_p(1),
\end{equation}
where $\bm \Sigma^{-1}=\mathrm{E}(-\partial^2 \ell(\bm{\theta}_0)/\partial \bm{\theta}\partial\bm{\theta}^\top)^{-1}$ and $o_p(1)\to 0$ in probability as $n\to\infty$.
Furthermore, it has been shown that the asymptotic distribution of $\sqrt{n}(\widehat{\bm{\theta}}_n-\bm{\theta}_0)$ is 
normal with zero mean $l\times 1$ vector and variance-covariance $l\times l$ matrix $\bm\Sigma^{-1}$ \citep[see][]{HALL1980155,Higgins1992,LingMcAleer2010}. 

%%==============================================================================================================%%
\section{\label{sec:new.test}{The proposed test statistics}}
%%==============================================================================================================%%

%\subsection{The hypotheses}
Let $k$ be the lag of the series with $k\in\{0, \pm 1, \pm 2,\cdots, \pm m\}$, where $m$ is the largest value considered for the auto-and-cross-correlations and define 
$$
\rho_{(r, s)}(\boldsymbol{\theta}_0,k) = \frac{\operatorname{Cov}\left(\varepsilon_{t}^{r}(\boldsymbol{\theta}_0), \varepsilon_{t-k}^{s}(\boldsymbol{\theta}_0)\right)}{\left\{\operatorname{Var}\left(\varepsilon_{t}^{r}(\boldsymbol{\theta}_0)\right) \operatorname{Var}\left(\varepsilon_{t}^{s}(\boldsymbol{\theta}_0)\right)\right\}^{1 / 2}} = \frac{\gamma_{(r,s)}(k)}{\sqrt{\gamma_{(r,r)}(0)}\sqrt{\gamma_{(s,s)}(0)}}\quad (r, s=1,2)
$$ 
as the lag- $k$ theoretical autocorrelation of the error process $\left\{\varepsilon_{t}(\boldsymbol{\theta}_0)\right\}$ where $\boldsymbol{\theta}_0$ is the true but unknown parameter vector.
Let
$$
\boldsymbol{\rho}(\boldsymbol{\theta}_0,k)=\left[\rho_{(r, r)}(\boldsymbol{\theta}_0,k), \rho_{(s, s)}(\boldsymbol{\theta}_0,k), \rho_{(r, s)}(\boldsymbol{\theta}_0,k)\right]^\top,
$$
and
$$
\mathbf{R}({\boldsymbol{\theta}_0})=\left[\mathbf{R}_{(r, r)}^\top({\boldsymbol{\theta}_0}), \mathbf{R}_{(s, s)}^\top({\boldsymbol{\theta}_0}), \mathbf{R}_{(r, s)}^\top({\boldsymbol{\theta}_0})\right]_{3 m \times 1}^\top 
$$
with
$$\mathbf{R}_{(r, s)}({\boldsymbol{\theta}_0})=\left[\rho_{(r, s)}({\boldsymbol{\theta}_0},1), \rho_{(r, s)}({\boldsymbol{\theta}_0},2), \ldots, \rho_{(r, s)}({\boldsymbol{\theta}_0},m)\right]^\top.
$$
We derive the asymptotic distribution of the proposed test statistics under the null hypothesis that the time series model in (\ref{AR.ARCH.model}) takes the correct functional forms given by $\mathbb{H}_0:\mu_t = \mu_t(\bm{\theta}_0)\text{ and }h_t = h_t(\bm{\theta}_0)$.
The alternative hypothesis is $\mathbb{H}_a:\mu_t \ne \mu_t(\bm{\theta}_0)\text{ or }h_t \ne h_t(\bm{\theta}_0)$.
Equivalently, the null and alternative hypotheses can be used for testing the lag residual auto-and-cross-correlation so that $\mathbb{H}_0: \mathbf{R}_{(r, s)}({\boldsymbol{\theta}_0})=\bm{0}_m$ and $\mathbb{H}_a: \mathbf{R}_{(r, s)}({\boldsymbol{\theta}_0}) \neq \bm{0}_m$, for all $r, s\in\{1,2\}$.
For simplicity, we dropped the symbol $\boldsymbol{\theta}_0$ so that $\rho_{(r, s)}({\boldsymbol{\theta}_0},k)=\rho_{(r, s)}(k)$ and $\mathbf{R}_{(r, s)}({\boldsymbol{\theta}_0}) = \mathbf{R}_{(r, s)}$.

Given a sample time series of length $n$ observations $z_1,z_2,\cdots,z_n$, under the assumptions of $\mathbb{H}_0$ and (\ref{eq:quasi}), we fit the model defined in \eqref{AR.ARCH.model}. Subsequently, we calculate the standardized residuals (conditional residuals) raised to powers $i\in{1,2}$ using the following expressions:
$$
\widehat{e}_{t}^{i}=\widehat{\varepsilon}_{t}^{i} \widehat{h}_{t}^{-i/2},
$$
where $\{\widehat{\varepsilon}_{t}\}, \{\widehat{\varepsilon}_{t}^2\}, \big\{\sqrt{\widehat{h}_{t}}\big\}$, and $\{\widehat{h}_{t}\}$ denote the 
sample residuals, squared-residuals, conditional volatility, and conditional variance of $z_t$, respectively. 

The corresponding sample correlation coefficient between the standardized residuals may be written as
\begin{equation}\label{CC.sample}
\widehat{r}_{(r,s)}(k)=\frac{\widehat{\gamma}_{(r,s)}(k)}{\sqrt{\widehat{\gamma}_{(r,r)}(0)}\sqrt{\widehat{\gamma}_{(s,s)}(0)}},
\end{equation}
where
${\widehat{\gamma}_{(r,s)}(k)=n^{-1}\sum_{t=k+1}^{n}(\widehat{e}_t^r-\widetilde{e}^r)(\widehat{e}_{t-k}^s-\widetilde{e}^s)}$,
for $k\ge 0$, $\widehat{\gamma}_{(r,s)}(-k)=\widehat{\gamma}_{(s,r)}(k)$, for $k<0$, is the autocovariance (cross-covariance), at lag-$k$, between the standardized residuals to {$r$th  power and the standardized residuals to $s$th power}, 
and $\widetilde{e}^i=n^{-1}\sum_{t=1}^{n}\widehat{e}_t^i$, for $i\in\{1,2\}$.

Under the regular assumptions, it can be shown that $\widetilde{e}=o_p(1)$, $\widetilde{e}^2=1+o_p(1)$, and {$n^{-1}\sum_{t=1}^{n}(\widehat{e}_{t}^2-\widetilde{e}^2)^2=\sigma^2+o_p(1)$,} where $\sigma^2$ converges to the value two
\citep[see][]{LiMak1994,Wong2005} and \citep[Theorem][]{LingMcAleer2003}. 
Hence, at lag-$k$, if we define $\bm \Gamma=(\bm \Gamma_{(r,r)},\bm \Gamma_{(s,s)},\bm \Gamma_{(r,s)})_{3m\times 1}^\top$ and  $\bm \Gamma_{(r,s)}=(\gamma_{(r,s)}(1),\cdots,\gamma_{(r,s)}(m))^\top$ as the
counterparts of 
$\widehat{\bm \Gamma}=(\widehat{\bm \Gamma}_{(r,r)},\widehat{\bm \Gamma}_{(s,s)},\widehat{\bm \Gamma}_{(r,s)})_{3m\times 1}^\top$ and
$\widehat{\bm \Gamma}_{(r,s)}=(\widehat{\gamma}_{(r,s)}(1),\cdots,\widehat{\gamma}_{(r,s)}(m))^\top$, respectively, with the replacement of the fitted residual $\widehat{\varepsilon}_t$ and conditional variance $\widehat{h}_t$ by $\varepsilon_t$ and $h_t$, respectively, we obtain:
\begin{eqnarray*}
	\widehat{\gamma}_{(1,1)}(k)&=&\frac{1}{n}\sum_{t=k+1}^{n}\frac{\widehat{\varepsilon}_{t}}{\sqrt{\widehat{h}}_{t}} \frac{\widehat{\varepsilon}_{t-k}}{\sqrt{\widehat{h}}_{t-k}},\quad
	\widehat{r}_{(1,1)}(k)=\frac{1}{n}\sum_{t=k+1}^{n}\frac{\widehat{\varepsilon}_{t}}{\sqrt{\widehat{h}}_{t}} \frac{\widehat{\varepsilon}_{t-k}}{\sqrt{\widehat{h}}_{t-k}},\\
	\widehat{\gamma}_{(2,2)}(k)&=&\frac{1}{n}\sum_{t=k+1}^{n}\left(\frac{\widehat{\varepsilon}_{t}^2}{\widehat{h}_{t}}-1\right) \left(\frac{\widehat{\varepsilon}_{t-k}^2}{\widehat{h}_{t-k}}-1\right),\quad
	\widehat{r}_{(2,2)}(k)=\frac{1}{n\sigma^2}\sum_{t=k+1}^{n}\left(\frac{\widehat{\varepsilon}_{t}^2}{\widehat{h}_{t}}-1\right) \left(\frac{\widehat{\varepsilon}_{t-k}^2}{\widehat{h}_{t-k}}-1\right),\\
	\widehat{\gamma}_{(1,2)}(k)&=&\frac{1}{n}\sum_{t=k+1}^{n}\frac{\widehat{\varepsilon}_{t}}{\sqrt{\widehat{h}_{t}}} \left(\frac{\widehat{\varepsilon}_{t-k}^2}{\widehat{h}_{t-k}}-1\right),\quad
	\widehat{r}_{(1,2)}(k)=\frac{1}{n\sigma}\sum_{t=k+1}^{n}\frac{\widehat{\varepsilon}_{t}}{\sqrt{\widehat{h}_{t}}} \left(\frac{\widehat{\varepsilon}_{t-k}^2}{\widehat{h}_{t-k}}-1\right),\\
	\widehat{\gamma}_{21}(k)&=&\frac{1}{n}\sum_{t=k+1}^{n}\left(\frac{\widehat{\varepsilon}_{t}^2}{\widehat{h}_{t}}-1\right)\frac{\widehat{\varepsilon}_{t-k}}{\sqrt{\widehat{h}_{t-k}}},\quad
	\widehat{r}_{(2,1)}(k)=\frac{1}{n\sigma}\sum_{t=k+1}^{n}\left(\frac{\widehat{\varepsilon}_{t}^2}{\widehat{h}_{t}}-1\right)\frac{\widehat{\varepsilon}_{t-k}}{\sqrt{\widehat{h}_{t-k}}}.
\end{eqnarray*}

We employ these autocorrelation coefficients to propose new portmanteau goodness-of-fit tests, as later defined in \eqref{new.test}, to check for linear and nonlinear dependencies within the residual series.

%%==============================================================================================================%%
%%==============================================================================================================%%
\begin{theorem}\label{theorem1}
Let the model defined in \eqref{AR.ARCH.model} be correctly specified and that (\ref{eq:quasi}) holds. 
Then, we have that
$$
\sqrt{n}({\widehat{\bm R}_{(1,1)}}^\top,{\widehat{\bm R}_{(2,2)}}^\top,{\widehat{\bm R}_{(r,s)}}^\top)^\top \overset{\rm D}{\to}\mathcal{N}_{3m}(\bm 0,\bm \Omega_{rs})\quad \text{as }n\to\infty,
$$
where 
\begin{equation}\label{R.corr}
\widehat{\bm R}_{(i,j)}=(\widehat{r}_{(i,j)}(1),\widehat{r}_{(i,j)}(2),\cdots,\widehat{r}_{(i,j)}(m))^\top,\quad \quad (i,j)\in\{(1,1),(2,2),(1,2),(2,1)\},
\end{equation}
$\overset{\rm D}{\to}$ denotes convergence in distribution and
$\bm \Omega_{rs} = \mathrm{E}\left[\bm{R}(\bm{\theta}_0)\bm{R}^\top(\bm{\theta}_0)\right]$ is the covariance matrix, which can be replaced by a consistent estimator $\widehat{\bm \Omega}_{rs}$: 
\begin{equation}\label{gamma.hat}
\widehat{\bm \Omega}_{rs}=
\left( 
\begin{array}{ccc}
\bm I_{m}-X_{11}\bm \Sigma^{-1}\bm X_{11}^\top& \bm 0& \bm 0\\
\bm 0& \bm I_{m}-\frac{1}{4}\bm X_{22}\bm \Sigma^{-1}\bm X_{22}^\top&  \bm 0\\
\bm 0&   \bm 0&\bm  I_{m}-\frac{1}{2}\bm X_{rs}\bm \Sigma^{-1}\bm X_{rs}^\top\\
\end{array}
\right),
\end{equation}
with $r\ne s\in \{1,2\}$, $\bm I_{m}$ denotes the identity $m\times m$ matrix,
\begin{equation}
\bm X_{11}(k)=\frac{1}{n}\sum_{t=k+1}^{n}\frac{1}{\sqrt{\widehat{h}_{t}}}\frac{\partial \mu_{t}}{\partial\bm{\theta}^\top} \frac{\widehat{\varepsilon}_{t-k}}{\sqrt{\widehat{h}_{t-k}}},
\end{equation}
\begin{equation}
\bm X_{22}(k)=\frac{1}{n}\sum_{t=k+1}^{n}\widehat{h}_{t}^{-1}\frac{\partial h_{t}}{\partial\bm{\theta}^\top}\left(\frac{\widehat{\varepsilon}_{t-k}^2}{\widehat{h}_{t-k}}-1\right),
\end{equation}
\begin{equation}
\bm X_{12}(k)=\frac{1}{n}\sum_{t=k+1}^{n}\frac{1}{\sqrt{\widehat{h}_{t}}}\frac{\partial \mu_{t}}{\partial\bm{\theta}^\top} \left(\frac{\widehat{\varepsilon}_{t-k}^2}{\widehat{h}_{t-k}}-1\right),
\end{equation}
and
\begin{equation}
\bm X_{21}(k)=\frac{1}{n}\sum_{t=k+1}^{n}\frac{1}{\sqrt{\widehat{h}_{t-k}}}\frac{\partial \mu_{t-k}}{\partial\bm{\theta}^\top} \left(\frac{\widehat{\varepsilon}_{t}^2}{\widehat{h}_{t}}-1\right).
\end{equation}
\end{theorem}
\begin{proof}
The proof is given in the Appendix (\ref{sec:Appendix}).
\end{proof}

By the results of Theorem \ref{theorem1}, we propose the portmanteau statistic, ${\dot{C}}_{rs}$ namely, and its modified version, $C_{rs}$ namely, to test $\mathbb{H}_0$ that the model stated in \eqref{AR.ARCH.model} is correctly specified. Thus, we have that
\begin{equation}\label{new.test}
{\dot{C}}_{rs}=n\left( 
\begin{array}{c}
\widehat{\bm R}_{(1,1)} \\
\widehat{\bm R}_{(2,2)}\\
\widehat{\bm R}_{(r,s)}
\end{array}
\right)^{\top}\widehat{\bm \Omega}_{rs}^{-1}
\left( 
\begin{array}{c}
\widehat{\bm R}_{(1,1)} \\
\widehat{\bm R}_{(2,2)}\\
\widehat{\bm R}_{(r,s)}
\end{array}
\right),\quad 
C_{rs}=n\left( 
\begin{array}{c}
\widetilde{\bm R}_{(1,1)} \\
\widetilde{\bm R}_{(2,2)}\\
\widetilde{\bm R}_{(r,s)}
\end{array}
\right)^{\top}\widehat{\bm \Omega}_{rs}^{-1}
\left( 
\begin{array}{c}
\widetilde{\bm R}_{(1,1)} \\
\widetilde{\bm R}_{(2,2)}\\
\widetilde{\bm R}_{(r,s)}
\end{array}
\right),
\end{equation}
where $\widehat{\bm R}_{(1,1)},\widehat{\bm R}_{(2,2)},\widehat{\bm R}_{(r,s)}$ are defined in \eqref{R.corr}, and
$\widetilde{\bm R}_{(1,1)},\widetilde{\bm R}_{(2,2)},\widetilde{\bm R}_{(r,s)}$ are obtained after replacing the autocorrelation coefficients in \eqref{CC.sample} by their standardized values formulated as
\begin{equation}\label{standardizedautocorrelation}
\displaystyle{\widetilde{r}_{(r,s)}(k)=\sqrt{\frac{n+2}{n-k}}\widehat{r}_{(r,s)}(k)}, \quad k\in \{1,\cdots, m\}.
\end{equation}
From the theorem on quadratic forms given in \cite{Box1954}, it is straightforward to show that ${\dot{C}}_{rs}$ and $C_{rs}$  are asymptotically chi-square distributed with $3m-(p+q+1)$ degrees of freedom.  

\begin{figure}[htb!]
\centering
\includegraphics[width=.85\textwidth,height=.35\textheight]{%figures/
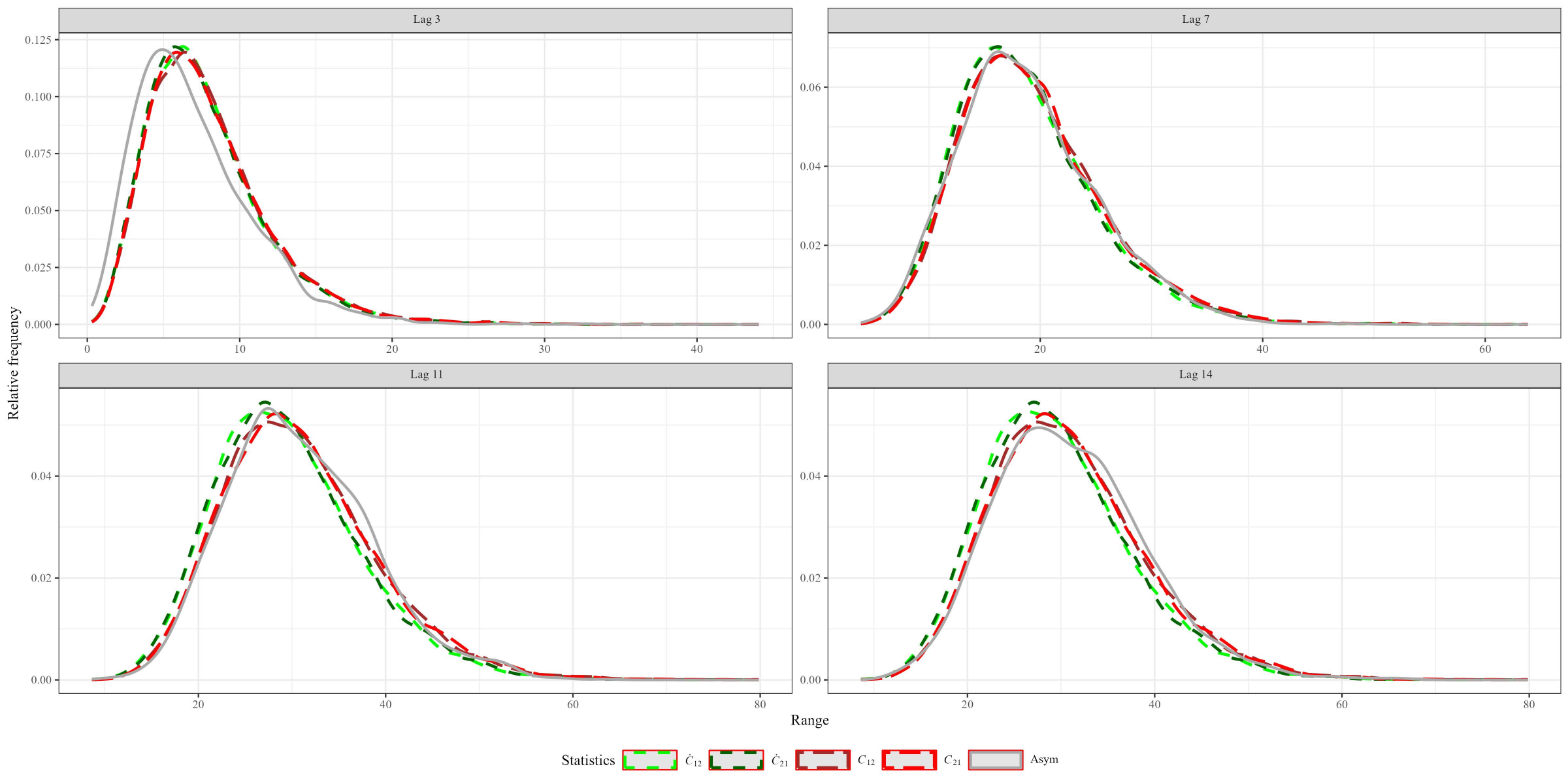}
\caption{The solid (dark-gray) line is the chi-square limiting distribution with degrees of freedom $3m-2$. The dot (green), dot (dark-green), longdash (brown), and longdash (red) are the Monte Carlo distributions of ${\dot{C}}_{12},{\dot{C}}_{21}, C_{12}$, and $C_{21}$, respectively, generated by 1000 replicates of a series of a length $n=200$ according to a Gaussian ARMA(1,1) model with $\phi_1=0.9$ and $\theta_1=-0.88$.
The tests are evaluated at each indicated lag $m\in\{3,7,11,14\}$.}
\label{fig:density}
\end{figure}

Figure \ref{fig:density} illustrates the accuracy of the approximation of the empirical distribution of ${\dot{C}}_{rs}$ and $C_{rs}$, for $r\ne s\in\{1,2\}$ to the chi-square distribution employing $10^3$ replicates when an ARMA(1,1) model fits to a sample of size $n=200$ generated from an ARMA(1,1) process defined as
\begin{equation}\label{eqnew:30}
z_t=0.9 z_{t-1}+\varepsilon_{t}-0.88\varepsilon_{t-1}.
\end{equation}
The parameters of the ARMA model stated in \eqref{eqnew:30} are selected to be very close to non-stationarity and non-invertibility case, whereas  the coefficient of the MA model is near to cancellation with the coefficient of the AR model to demonstrate the usefulness of the proposed tests, even with extreme cases.

We found similar results for small and large samples and our preliminary analysis indicates that the portmanteau tests based on the statistics $C_{rs}$ control the type I error probability more successfully than the tests that consider the statistics ${\dot{C}}_{rs}$. 
Hence, we recommend the use of $C_{rs}$.

\begin{remark}
As mentioned, the proposed test statistics in \eqref{new.test} can be seen as combinations of the statistics presented by \cite{Wong2005} and \cite{ZM2019}.  
Thus, each test statistic $C_{rs}$ may be seen as a linear combination of three existent test statistic proposed by \cite{LjungBox1978}, \cite{McLeodLi1983}, and \cite{ZM2019}, modifying the corresponding statistic ${\dot{C}}_{rs}$, which are linear combinations of three statistics given by \cite{BoxPierce1970}, \cite{LiMak1994}, and \cite{ZM2019}. 
\end{remark}

%%==============================================================================================================%%
\section{\label{sec:simulation}Simulation studies}
%%==============================================================================================================%%

We carry out Monte Carlo simulations  to examine statistical properties of the proposed tests.
For comparative purposes, we also consider three test statistics given by
\begin{equation}\label{McLeodLitest}
Q_{rs}=n(n+2)\sum_{k=1}^{m}(n-k)^{-1}\widehat{r}_{(r,s)}^2(k),\quad (r,s)\in\{(1,2),(2,1),(2,2)\},
\end{equation}
where $Q_{12}$ and $Q_{21}$ represent the statistics presented in \cite{ZM2019} and
$Q_{22}$ denotes the statistic proposed by \cite{McLeodLi1983}.
In addition, we consider two statistics introduced by \cite{LiMak1994} and \cite{Wong2005}, which are denoted by $Q_{\textrm{LM}}$ and $Q_{\textrm{WL}}$, respectively, given by
\begin{eqnarray}\label{LiMaktest}
Q_{\textrm{LM}}&=&n\sum_{k=1}^{m}\widehat{r}_{(2,2)}^2(k),\\
\label{WongLingtest}
Q_{\textrm{WL}}&=&n\left( 
\begin{array}{c}
\widehat{\bm R}_{(1,1)} \\
\widehat{\bm R}_{(2,2)}
\end{array}
\right)^{\top}
\left( 
\begin{array}{cc}
\bm I_{m}&  \bm 0 \\
\bm 0 & \bm I_{m}-\frac{1}{4}\bm X_{22}\bm \Sigma^{-1}\bm X_{22}^\top
\end{array}
\right)
^{-1}
\left( 
\begin{array}{c}
\widehat{\bm R}_{(1,1)} \\
\widehat{\bm R}_{(2,2)}
\end{array}
\right).
\end{eqnarray}
First, we examined six statistics, $C_{12}, C_{21},Q_{12},Q_{21}$, $Q_{22}$ and $Q_{\textrm{WL}}$ namely, assuming the following five linear models studied by \cite{ZM2019}:
%\clearpage
\begin{description}
\item [\rm {\bf A1.} AR(1) model:] $z_t=-0.9z_{t-1}+\varepsilon_t$;
\item [\rm {\bf A2.} AR(2) model:] $z_t=0.6z_{t-1}-0.5z_{t-2}+\varepsilon_t$;
\item [\rm {\bf A3.} MA(1) model:] $z_t=0.8\varepsilon_{t-1}+\varepsilon_t$;
\item [\rm {\bf A4.} ARMA(2,1) model:] $z_t=0.8z_{t-1} +0.15z_{t-2}+0.3\varepsilon_{t-1}+\varepsilon_t$;
\item [\rm {\bf A5.} ARMA(1,1) model:] $z_t=0.6z_{t-1} +0.4\varepsilon_{t-1}+\varepsilon_t$.
\end{description}
For model A1, the parameter used by \cite{ZM2019} was $\phi=0.8$. However, we considered here a value negative of the parameter close to non-stationarity case to assess the behavior of the test statistics associated with positive and negative values of the parameters. We also analyzed for this model the cases where $\phi\in \{\pm 0.1, \pm 0.5, \pm 0.8\}$ whose results showed very minor changes in the behavior of the proposed tests. For model A3, we also explored the cases where ${\theta}\in \{\pm 0.3, \pm 0.6, \pm 0.9\}$ and obtained good results. 

Second, we investigated four statistics, $C_{12},C_{21},Q_{\textrm{WL}}$, and $Q_{\textrm{LM}}$ namely, according to the following three nonlinear models studied by \cite{Velasco2015} and \cite{NGAI2017}:
\begin{description}
\item [\rm {\bf A6.} GARCH(1,1) model:] $\varepsilon_t=\xi_t\sqrt{h_t},
\, h_t=0.1+0.3\varepsilon_{t-1}^2+0.5 h_{t-1}$;
\item [\rm {\bf A7.} AR(1)-ARCH(1) model:] $z_t=0.5z_{t-1}+\varepsilon_t,
\,\varepsilon_t=\xi_t\sqrt{h_t},
\, h_t=0.1+0.4\varepsilon_{t-1}^2$;
\item [\rm {\bf A8.} AR(1)-GARCH(1,1) model:] $z_t=0.5z_{t-1}+\varepsilon_t,
\, \varepsilon_t=\xi_t\sqrt{h_t}, h_t=0.1+0.3\varepsilon_{t-1}^2+0.5 h_{t-1}$;
\end{description}
where $\eta_t\overset{\mathrm{i.i.d.}}{\sim}\mathcal{N}(0,1)$ or Student-$t$ distribution with 10 degrees of freedom.

In all experiments, we use the \texttt {R} software \citep[\url{www.R-project.org},][]{R2020} to simulate 1000 replicates of artificial series of size
$n+n/2$ with $n\in\{100,300\}$. 
However, only the last $n$ data points are used to carry out portmanteau tests with the residuals of some fitted models. 

The empirical size and power of the tests are calculated based on a nominal level 5\%.
Simulation results for nominal levels 1\% and 10\% are not reported, due to space conservation, but they are available upon request.

First, we calculate the type I error probability, at lags $m \in \{5,10\}$, based on six statistics, $C_{12}, C_{21},Q_{12},Q_{21}$, $Q_{22}$ and $Q_{\textrm{WL}}$ namely, when a true model is fitted to a series generated according to models A1-A5. 
Second, we investigated the accuracy of estimating type I error probability using four statistics, $C_{12},C_{21},Q_{\textrm{WL}}$, and $Q_{\textrm{LM}}$ namely, when a true model is fitted to a series generated according to models A6-A8. 

\begin{table}[H]
\small
\centering
\renewcommand{\arraystretch}{0.6}
\caption{\label{tab:table1}\small Empirical sizes, for 5\% significance test, of the indicated statistic, distribution, model, $n$, and $m$.}
\begin{tabular}{lc cc cccccc cccccc}
\toprule
\multirow{2}{*}{Model}&\multirow{2}{*}{$n$} &&\multicolumn{6}{c}{$m=5$ }&&\multicolumn{6}{c}{$m=10$}\\
\cline{4-9}\cline{11-16}
&&& $C_{12}$&$C_{21}$&$Q_{12}$& $Q_{21}$ &$Q_{22}$&$Q_{\textrm{WL}}$  &&  $C_{12}$&$C_{21}$&$Q_{12}$& $Q_{21}$ &$Q_{22}$ &$Q_{\textrm{WL}}$ \\
\midrule
\multicolumn{16}{c}{Gaussian distribution}\\ 
\midrule
A1 &$100$&& 5.7 & 5.8 & 4.4 & 4.4 & 4.4 & 3.6 && 5.7 & 5.9 &  4.7 & 4.7 & 4.7 & 4.4\\
&$300$&&5.6 & 5.4 & 4.9 & 4.9 & 5.0 & 3.7 && 5.6 & 5.5 & 4.9 & 4.9 & 5.1 & 4.3 \\
A2 &$100$&&5.2 & 4.8 & 4.2 & 4.2 & 4.1 & 3.1 && 5.4 & 4.9 & 4.3 & 4.3 & 4.6 & 3.6 \\
&$300$&&5.2 & 5.5 & 5.0 & 5.0 & 5.0 & 3.5  && 5.4 & 5.3 & 4.8 & 4.8 & 5.2 & 4.0 \\
A3 &$100$&&6.0 & 5.8 & 4.4 & 4.4 & 4.3 & 3.7 && 6.0 & 6.0 & 4.6 & 4.6 & 4.8 & 4.4 \\
&$300$&& 5.5 & 5.8 & 5.1 & 5.1 & 5.0 & 3.7&& 5.8 & 5.8 & 5.0 & 5.0 & 5.1 & 4.3\\
A4 &$100$&& 6.6  & 6.2 & 4.3 & 4.3 & 3.9 &  4.1  && 6.2  & 5.9 & 4.3  &  4.4 &  4.5 & 4.2\\
&$300$&& 6.5  & 6.4 & 4.5 & 4.7 & 4.6 &  4.5  &&  5.6  & 5.2 & 5.4 & 4.2 & 4.8 &  4.3\\
A5 &$100$&& 5.5 & 5.3 & 4.3 & 4.4 & 4.0 & 2.9   &&  5.6 & 5.4 & 4.3  & 4.3  &  5.2 & 4.0\\
&$300$&&5.3 & 5.6 & 4.8 & 4.8 & 5.0 & 3.6  && 5.5 & 5.6 & 4.8 & 4.8 & 5.1 & 4.0\\
\midrule
\multicolumn{16}{c}{Student-$t$ distribution with 10 degrees of freedom}\\ 
\midrule
A1 &$100$&&5.8 & 5.4 & 4.1 & 4.1 & 4.3 & 3.4 && 5.6 & 5.1 & 4.0 & 4.0 & 4.3 & 3.9 \\
&$300$&&6.2 & 6.3& 4.5 & 4.5 & 5.3 & 4.2  && 6.1 & 6.3 & 5.0 & 5.0 & 5.3 & 4.6 \\
A2 &$100$&&4.5 & 4.6 & 3.9 & 3.9 & 3.9 & 2.9 && 4.6 & 4.5 & 3.6 & 3.6 & 3.9 & 3.3\\
&$300$&& 5.7 & 5.7 & 4.3 & 4.3 & 5.2 & 3.5 && 5.8 & 5.8 & 4.7 & 4.7 & 5.5 & 4.0\\
A3 &$100$&& 5.6 & 5.2 & 4.4 & 4.4 & 4.3 & 3.4 && 5.4 & 5.3 & 4.1 & 4.1 & 4.3 & 3.8\\
&$300$&& 5.9 & 6.0 & 4.5 & 4.5 & 5.2 & 4.0  && 6.0 & 6.4 & 5.0 & 5.0 & 5.5 & 4.4 \\
A4 &$100$&& 7.2  & 7.4 & 4.9 & 5.1 & 4.2 & 4.0   && 5.8  & 6.0 &  4.8 &  5.0 & 4.7  & 4.2\\
&$300$&& 6.2  & 6.0 & 4.6 & 4.8 & 5.0 & 4.4   && 5.8  & 5.6 &  4.8 &  5.0 & 5.2  & 4.4\\
A5 &$100$&& 5.0 & 4.7 & 4.2 & 4.2 & 4.1 & 2.9 && 4.9 & 4.8 & 3.9 & 3.9 & 4.1 & 3.6 \\
&$300$&& 5.6 & 5.9 & 4.3 & 4.3 & 5.1 & 3.8  && 5.7 & 5.8 & 4.7 & 4.7 & 5.5 & 4.1 \\
\bottomrule
\end{tabular}
\end{table}

The empirical sizes of the test corresponding to the nominal size $5\%$ over 1000 independent simulations belong to the 95\% confidence interval $[3.65\%,6.35\%]$ and to the 99\% confidence interval $[3.22\%, 6.78\%]$. 
From Tables \ref{tab:table1} and \ref{tab:table2}, note that tests based on the statistics $Q_{\textrm{WL}}$  and $Q_{\textrm{LM}}$ can distort the test size, whereas the tests using the statistics $Q_{22},Q_{12},Q_{21}$ and the proposed statistics exhibit no substantial size distortion and, generally, have empirical levels that improve as $n$ increases. Also, we found similar results for the cases where the error terms have either skew-normal distribution with asymmetry parameter $\kappa\in\{-1.0, -0.5, 0.5, 1.0\}$ or Student-$t$ distribution with degrees of freedom $\nu \in\{5,15,20\}$. These results are omitted here due to space conservation, but they are available upon request. 

\begin{table}[H]
\small
\centering
\renewcommand{\arraystretch}{0.7}
\caption{\label{tab:table2}Empirical sizes, for 5\% significance test, of the indicated statistic, distribution, model, $n$, and $m$.}
\begin{tabular}{cccc cccc cccc}
\toprule
\multirow{2}{*}{Model}&\multirow{2}{*}{$n$} &&\multicolumn{4}{c}{$m=5$ }&&\multicolumn{4}{c}{$m=10$}\\
\cline{4-7}\cline{9-12}
&&& $C_{12}$&$C_{21}$&$Q_{\textrm{WL}}$&$Q_{\textrm{LM}}$  &&  $C_{12}$&$C_{21}$&$Q_{\textrm{WL}}$ &$Q_{\textrm{LM}}$\\
\midrule
\multicolumn{12}{c}{Gaussian distribution}\\ 
\midrule
A6 &$100$&& 3.7 &  3.8 &   2.3& 2.8  &&  4.1  & 4.0 & 2.7 & 2.3\\
&$300$&& 3.5 & 3.4 & 1.8 & 3.2    && 3.5  & 3.6  & 2.5 & 2.5\\
A7 &$100$&& 3.9 & 3.9 & 2.3 & 3.5   && 4.5 & 45 & 3.2 &3.1\\
&$300$&& 3.8 & 3.7 &  2.1 & 3.6  &&  4.5 & 4.4  & 3.2 &4.1\\
A8 &$100$&& 4.0 & 3.8 &  2.3 &  2.8 && 4.0  &  3.8 & 2.7 &2.4\\
&$300$&& 3.4 & 3.3 & 1.8 &  3.0 &&  3.3 &  3.4 & 2.3 & 2.3\\
\midrule
\multicolumn{12}{c}{Student-$t$ distribution with 10 degrees of freedom}\\ 
\midrule
A6 &$100$&&  4.4& 4.0 & 2.4& 2.8 && 4.2  & 3.7  & 2.6 &2.1\\
&$300$&& 4.4 & 4.1 & 2.9 & 4.1 && 4.2  &  4.2 &2.9 &3.0\\
A7 &$100$&& 4.4 & 4.3 & 2.6& 3.0 && 4.8  & 4.7  & 3.4 &2.0\\
&$300$&& 5.1 & 5.0 &  3.3& 4.8 &&  5.1 &  5.2 & 3.7 &3.3\\
A8 &$100$ && 4.8&  4.3 &  2.7& 2.7  &&  4.3 &  4.0 & 2.9 &1.4\\
&$300$&&4.7  & 4.8 &  2.9& 3.6 && 4.1  &  4.4 & 2.9 &2.7\\
\bottomrule
\end{tabular}
\end{table}

%%==============================================================================================================%%
\subsection{\label{sec:power1}Testing linearity in linear time series models}
%%==============================================================================================================%%

Now, we investigate the efficiency to distinguish power for the mean term of the test statistics $C_{12},C_{21},Q_{12},Q_{21},Q_{22}$, and $Q_{\textrm{WL}}$\footnote{We also test the form of the GARCH-type models by examining the GARCH(1,1) model versus the MA(1)-GARCH(1,1)
and AR(1)-GARCH(1,1) models based on the test statistics $C^\star,Q_{\textrm{WL}}$, and $Q_{\textrm{LM}}$.
The results are available upon request}.
For expositional simplicity, we define $C^\star$ as the highest test power attained by the statistics $C_{12}$ and $C_{21}$, whereas $Q^\star$ as the highest test power attained by $Q_{12}$ and $Q_{21}$,  that is, 
\begin{equation}\label{cstar}
C^\star=\max(C_{12},C_{21}),\quad Q^\star=\max(Q_{12},Q_{21}).
\end{equation}
The power of the tests are calculated under the null hypothesis $\mathbb{H}_0$ that $z_t$ satisfies the ARMA model 
$$
z_t = \mu_t + \sum_{i =1}^{p}\phi_i z_{t-i} + \sum_{i =1}^{q}\theta_i \varepsilon_{t-i} + \varepsilon_t,
$$
which can be seen as an AR($p$) model, where $p\to\infty$.
We followed the approach presented by \cite{Ng2005} who used the Bayesian information criterion (BIC) to select the order $p\in\{0,1,\cdots,\lfloor 8(n/100)^{1/4}\rfloor\}$, where $\lfloor a \rfloor$ denotes the floor function (integer part) of the number $a\in\mathbb{R}$, when an AR($p$) model erroneously fits to series generated from the following models studied by \cite{LiMak1994}, \cite{Wong2005}, \cite{NGAI2017}, and \cite{ZM2019}:\\[-0.8cm]
\begin{description}
\item [\rm {\bf B1.} Bilinear (BL) model:] $z_t=0.2+0.4z_{t-1}+\varepsilon_{t}+\varphi z_{t-1}\varepsilon_{t-1}$, where $\varepsilon_t\overset{\mathrm{i.i.d.}}{\sim}\mathcal{N}(0,1)$, with parameter values of $\varphi$ being selected in the range $0<\varphi<2.5$;\\[-0.8cm]
\item [\rm {\bf B2.} Random coefficient AR (RCAR) model:] $z_t= 0.2z_{t-1}+u_{t}, u_{t}=\varphi\eta_{t} z_{t-1}+\varepsilon_{t}$, where $\{\varepsilon_t\}$ and $\{\eta_t\}$ are two sequences of i.i.d.\,$\mathcal{N}(0,1)$ random variables, which are independent from each other variable; note that the RCAR model is a special case of the AR(1) and ARCH(1) models as observed in
$\mathrm{E}(u_t^2|\mathcal{F}_{t-1})=\varphi^2 z_{t-1}^2+1$, which is the conditional variance 
over time. 
We select parameter values of $\varphi$ from the range $0<\varphi<2.5$;\\[-0.8cm]
\item [\rm {\bf B3.} TAR model:] $z_t=0.8 z_{t-1}\mathbb{I}_{\{z_{t-1}\le -1\}} - 0.8 z_{t-1}\mathbb{I}_{\{z_{t-1}> -1\}}+\varepsilon_t$, where $\varepsilon_t\overset{\mathrm{i.i.d.}}{\sim}\mathcal{N}(0,1)$;\\[-0.8cm]
\item [\rm {\bf B4.} AR(1)-ARCH(2) model:] $z_t=0.2 z_{t-1}+\varepsilon_{t},
\quad
\varepsilon_t=\xi_t\sqrt{h_t}, 
\quad h_t=0.2+0.2\varepsilon_{t-1}^2+0.2\varepsilon_{t-2}^2$, 
where $\xi_t\overset{\mathrm{i.i.d.}}{\sim}\mathcal{N}(0,1)$.
\end{description}
Note that we found similar results based on the Akaike information criterion --AIC-- \citep{Akaike1974}.

\begin{figure}[htb!]
\centering
\includegraphics[width=0.9\textwidth,height=0.35\textheight]{%figures/
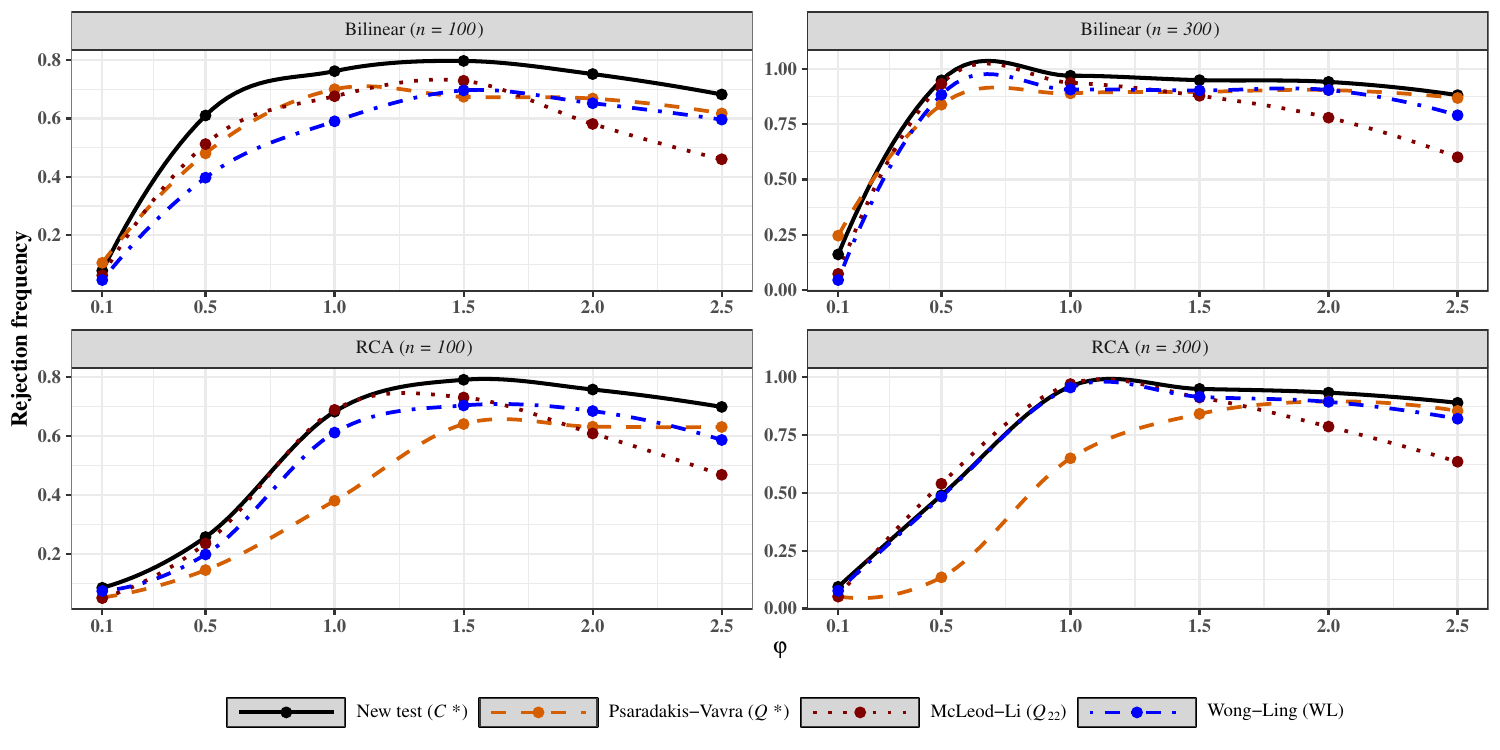}
\caption{Rejection frequencies of the indicated statistic for the bilinear (B1) and RCAR (B2) models with $0<\varphi<2.5$ and listed sample sizes $n$.}
\label{fig:B1.2}
\end{figure}

\begin{figure}[htb!]
\centering
\includegraphics[width=0.9\textwidth,height=0.375\textheight]{%figures/
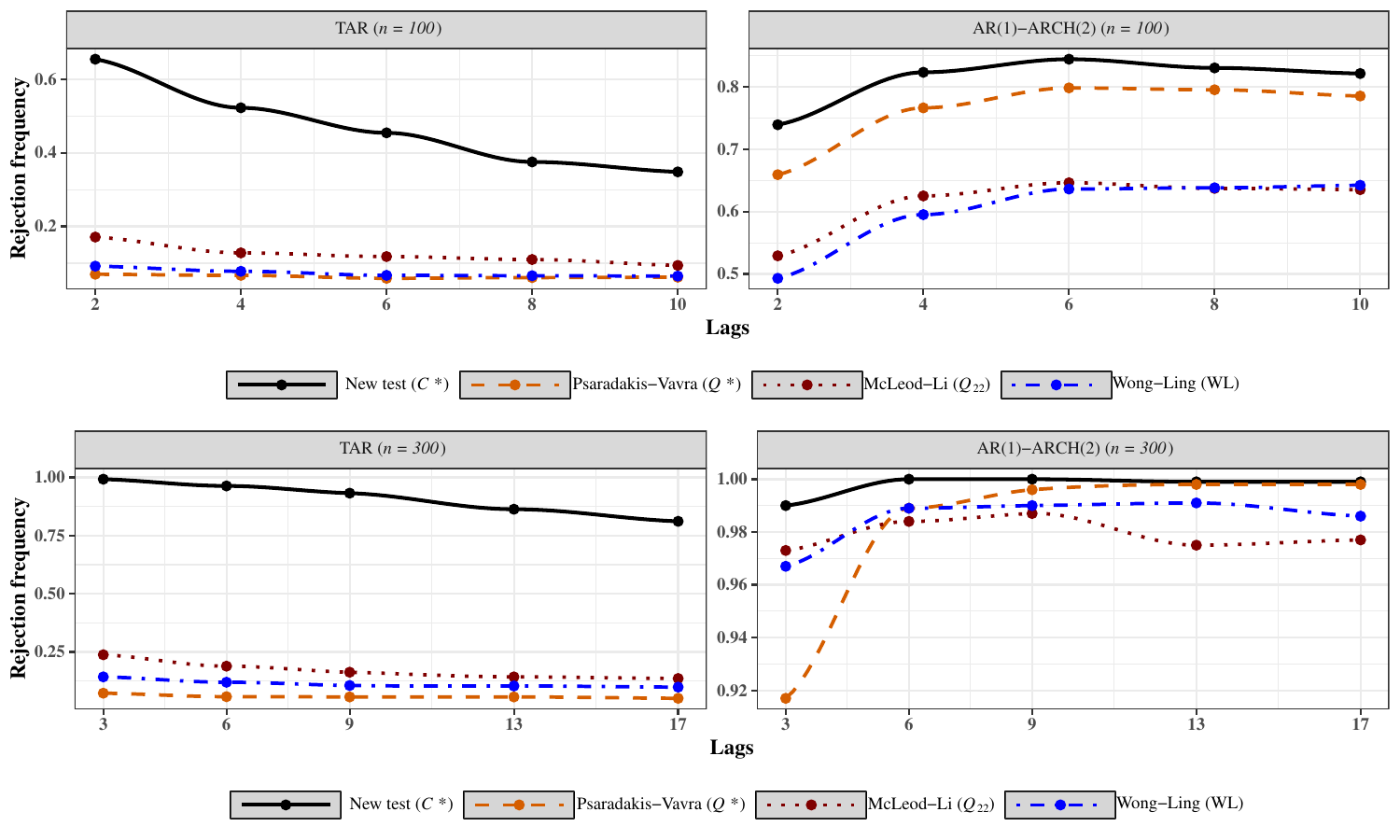}
\vspace{-0.35cm}
\caption{Rejection frequencies of the indicated statistic for the TAR (B3) and AR(1)-ARCH(2) (B4) models with the listed sample size $n$.}
\label{fig:B3.4}
\end{figure}

Figure \ref{fig:B1.2} displays the rejection frequencies considering a 5\% nominal level of the statistics $C^\star,Q^\star,Q_{22}$, and $Q_{\textrm{WL}}$ when an AR($p$) model erroneously fits data of size $n\in\{100,300\}$ from BL (B1) and RCAR (B2) models
at lag value $m=\lfloor \sqrt{n}\rfloor$.
The results for both models are based on parameter values $\varphi$ varying from 0 to 2.5, as mentioned.
In addition, Figure \ref{fig:B3.4} shows the rejection probability of the aforementioned tests (of nominal level $0.05$) employing the nonlinear TAR (B3) and AR(1)-ARCH(2) (B4) models.
For $n\in\{100,300\}$, the tests are calculated at lags $m\in\{2,4,6,8,10\}$ and $m\in\{3,6,9,13,17\}$, respectively.  
From Figures \ref{fig:B1.2} and \ref{fig:B3.4}, note that the performance of the proposed statistic $C^\star$ is, in general, the best, especially for small sample sizes.
Thus, we conclude that the proposed statistics are helpful for testing for linearity of stationary time
series.

%%==============================================================================================================%%
\subsection[Testing the AR-ARCH models]{Testing the AR-ARCH models}
\label{models:ararch}
%%==============================================================================================================%%

In order to examine the ability for discriminating power for the mean and conditional variance parts of a time series model, we consider the AR-GARCH model versus 
nonlinear models with GARCH errors.
The process $\{z_t\}$ satisfies the null hypothesis $\mathbb{H}_0$ of heteroskedasticity given by
$$
z_t=\phi z_{t-1}+\varepsilon_t,
\quad
\varepsilon_t=\xi_{t}\sqrt{h_{t}},
\quad
h_t=\omega+\alpha_1\varepsilon_{t-1}^2,
\quad \xi_t\overset{\mathrm{i.i.d.}}{\sim}\mathcal{N}(0,1).$$
The alternative models are:
\begin{description}
\item [\rm {\bf D1.} AR(1)-ARCH(2) model:] $z_t=0.5z_{t-1}+ \xi_{t}\sqrt{h_{t}},
\quad h_t=0.01+0.4\varepsilon_{t-1}^2+0.3\varepsilon_{t-2}^2$;
\item [\rm {\bf D2.} AR(1)-GARCH(1,1) model:] $z_t=0.5z_{t-1}+ \xi_{t}\sqrt{h_{t}},
\quad
h_t=0.01+0.4\varepsilon_{t-1}^2+0.5 h_{t-1}$;
\item [\rm {\bf D3.} AR(2)-ARCH(2) model:] $z_t=0.5z_{t-1}+0.2z_{t-2}+ \xi_{t}\sqrt{h_{t}},
\quad
h_t=0.01+0.4\varepsilon_{t-1}^2+0.2 \varepsilon_{t-2}^2$;
\item [\rm {\bf D4.} TAR model with GJR-GARCH(1,1) error:] 
$$
z_t=0.4z_{t-1}+0.5z_{t-1}\mathbb{I}_{\{z_{t-1}>0\}}+\xi_{t}\sqrt{h_{t}},
\quad 
h_t=0.1+(0.3+0.4 \mathbb{I}_{\{\varepsilon_{t-1}<0\}})\varepsilon_{t-1}^2+0.4 h_{t-1}.$$
\end{description}

\begin{figure}[htb!]
\centering
\includegraphics[width=16cm, height=10cm]{%figures/
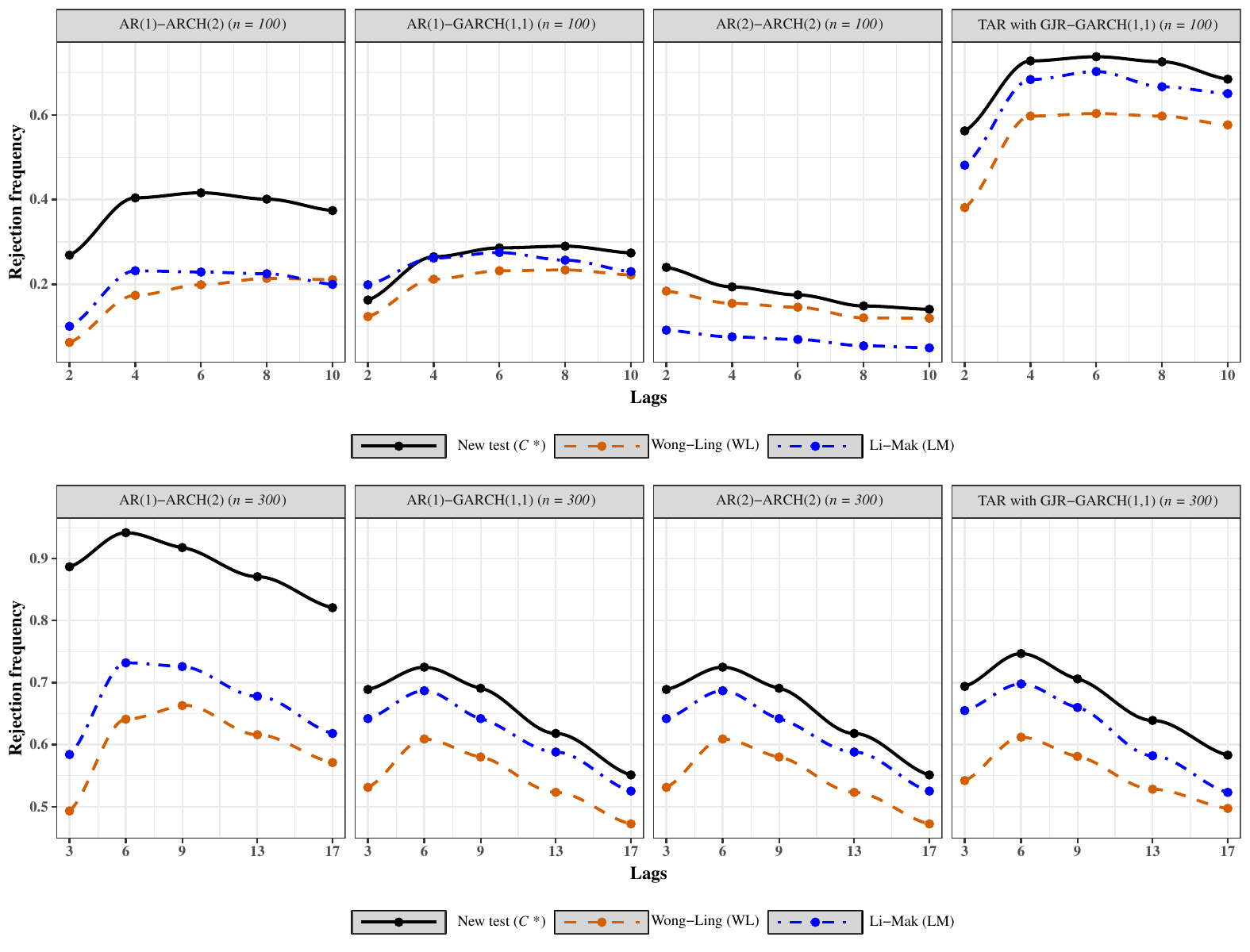} 
\caption{Empirical power based on 5\% nominal level of the indicated statistic and sample size $n$ when AR(1)-ARCH(1) erroneously fits data generated from AR(1)-ARCH(2), AR(1)-GARCH(1,1), AR(2)-ARCH(2), and TAR with GJR-GARCH(1,1) models named as D1, D2, D3, and D4, respectively.}
\label{fig:D.1.4}
\end{figure}

For each model, the power of the test for the statistics $C^\star,Q_{\textrm{WL}}$, and $Q_{\textrm{LM}}$ are calculated at lags $m\in\{2,4,6,8,10\}$ and $m\in\{3,6,9,13,17\}$ associated with $n\in\{100,300\}$. The results are shown in Figure \ref{fig:D.1.4}. 
Note that the performance of the proposed test is in general better when compared with the other two tests. 

Worth noting that, in general, as the lag order increases, the power of the portmanteau tests decreases, especially under models like the TAR model, for several reasons including the following:
\begin{itemize}
\item 	When the lag is large compared to the overall sample size, we usually get less reliable estimates of autocorrelation and, consequently, less power in the portmanteau test.
\item As we increase the lag order, we are including more lags in the test, which means we are estimating more parameters. This results in a loss of degrees of freedom in the test statistics. With fewer degrees of freedom, the test becomes less sensitive to detecting the absence of autocorrelation.
\item TAR models have different nonlinear thresholds which complicate the estimation of the autocorrelation function at higher lag orders, making it more challenging to detect the absence of autocorrelation.
\end{itemize}

%%==============================================================================================================%%
\section{\label{sec:application}Empirical applications}
%%==============================================================================================================%%

%%==============================================================================================================%%
\subsection{\label{sec:application1}Test for nonlinearity in AR models using stock returns}
%%==============================================================================================================%%

We demonstrate the usefulness of the proposed tests for detecting nonlinearity in AR models for a set of weekly stock returns.
We select 92 companies studied by \cite{Kapetanios2009TestingFS} and \cite{ZM2019}.
These companies are a subset of the Standard \& Poor 500 composite index (S\&P 500), spanning over the period from 18 June  1993 to 31 December 2007 ($n=781$ observations). 

Following the procedure presented by \cite{ZM2019}, we fit an AR($p$) model for each series, where the order of $p$ was selected by minimizing the BIC according to the algorithm explained in Section \ref{sec:power1} \citep{Ng2005}.
The asymptotic p-values (at 5\% significance level) for tests based on the statistics $C^\star, Q^\star, Q_{22}$, and $Q_{\textrm{WL}}$, with $m=\ln\lfloor n\rfloor$, are reported in Table \ref{tab:table4}.
Since the conditional heteroskedasticity is often considered as a main characteristic of asset returns, we expect that the AR model will not capture the nonlinear features in most of the stock returns considered in our analysis and the null hypothesis of linearity should be rejected.

\begin{table}[htbp]
\caption{\label{tab:table4}p-values for testing the neglected nonlinearity in AR models with the indicated statistic for the listed company from the S\&P 500 index.}
%\scriptsize
\small
\centering
\renewcommand{\arraystretch}{0.5}
\renewcommand{\tabcolsep}{0.1cm}
\begin{tabular}{lllllllllll}
\toprule
\multirow{2}{*}{Company}&\multicolumn{4}{c}{p-value} &&\multirow{2}{*}{Company}&\multicolumn{4}{c}{p-value} \\
%\cline{2-5}\cline{8-11}
& $C^\star$ & $Q^\star$ & $Q_{22}$ &$Q_{\textrm{WL}}$&&& $C^\star$ & $Q^\star$ & $Q_{22}$ &$Q_{\textrm{WL}}$\\
\midrule
Alcoa Inc                 & 0.00  &0.00& 0.00 & 0.00         &&Deere and Co. & 0.00   & 0.03&   0.00  &  0.01 \\  
Apple Inc.                & 0.00  &0.01& 0.41  & 0.19        &&D.R. Horton & 0.06  &0.21&   0.02   &   0.11 \\ 
Adobe Systems Inc         & 0.00  &0.15 & 0.00   &   0.00    && Danaher Corp.& 0.00  &0.01& 0.00 & 0.00\\  
Analog Devices Inc        & 0.00  &0.00 & 0.00   &  0.00     &&Walt Disney Co.& 0.00      & 0.10  &0.00&0.01\\
Archer-Daniels-Midland    & 0.07  &0.52&   0.01   & 0.05     &&Duke Energy& 0.00  &0.00& 0.00 & 0.00\\
Autodesk Inc              & 0.00  &0.06&  0.01    &   0.02   &&Ecolab Inc.& 0.00  &0.00& 0.00 & 0.00\\
American Electric Power   &  0.00 & 0.00& 0.00     &   0.00  &&Equifax Inc.&  0.96    & 0.55 &0.99&0.99\\
AES Corp                  &  0.00 & 0.00&  0.00    &   0.00  && Edison Int’l&0.00  &0.00& 0.00 & 0.00\\        
AFLAC Inc                 &   0.00 & 0.00&  0.00  &  0.00    &&EMC Corp.&  0.01    & 0.02 &0.03&0.12\\           
Allergan Inc             &  0.95 &0.46&  0.92  &    0.99     &&Emerson Electric&0.00  &0.03& 0.00 & 0.00\\
American Intl Group Inc  &  0.00 &0.00&  0.00  &    0.00     &&Equity Residential&0.00  &0.41& 0.00 & 0.00\\          
Aon plc                  & 0.00  &0.00&   0.00   & 0.00      &&EQT Corporation& 0.00  &0.03& 0.00 & 0.00  \\
Apache Corporation       & 0.00 &0.02&   0.00   &  0.00      &&Eaton Corp.&   0.04   & 0.21 & 0.51&0.54\\
Anadarko Petroleum       &  0.00 & 0.15& 0.00     &  0.00    &&Entergy Corp.&0.00  &0.03& 0.00 & 0.00\\
Avon Products            &  0.00 & 0.17&  0.09    &  0.01    &&Exelon Corp.&  0.22    &  0.33 &0.06&0.29\\
Avery Dennison Corp      &  0.00 &0.00&   0.00   & 0.00      &&Ford Motor&0.00  &0.04& 0.00 & 0.01\\
American Express Co      &  0.00 &0.00&   0.00   &    0.01   && Fastenal Co&  0.09    & 0.14 &0.04 & 0.22\\
Bank of America Corp     &   0.00 & 0.00&    0.00   &   0.00 &&FedEx Corporation&0.00  &0.03& 0.00 & 0.00 \\
Baxter International Inc. &  0.00 & 0.00&    0.33  &  0.20   &&Fiserv Inc&0.00  &0.00& 0.00 & 0.00\\          
BBT Corporation           & 0.00  &0.01&  0.00    &  0.00    &&Fifth Third Bancorp& 0.00  &0.00& 0.00 & 0.00 \\          
Best Buy Co. Inc.         & 0.00  &0.01&   0.00   &  0.00    &&Fluor Corp.&  0.01    &   0.03 &0.01&0.05\\          
Becton Dickinson          & 0.00  &0.10 &  0.00    &   0.00  &&Frontier Commun.& 0.00  &0.00& 0.00 & 0.00 \\
Franklin Resources        &  0.00 &0.00&    0.00  &  0.00    &&Gannett Co.&0.00  &0.09& 0.02  & 0.02\\
Brown-Forman Corp         & 0.00   &0.00 & 0.02   &   0.00   &&General Dynamics& 0.00     & 0.06 &0.00&0.02\\
Baker Hughes Inc          &  0.00  &0.00 &  0.00  &  0.00    &&General Electric&0.00  &0.00& 0.00 & 0.00 \\
The Bank of NY Mellon     &  0.00  &0.00 &  0.00  &  0.00    &&General Mills& 0.54     & 0.77 &0.81& 0.41\\
Ball Corp                 &  0.00  &0.00 &  0.00  &  0.00    &&Genuine Parts&  0.00 &0.01&   0.00   & 0.00  \\
Boston Scientific         & 0.01  & 0.04&   0.01   &  0.12   &&Gap (The)& 0.00  &0.00&   0.00   & 0.00   \\
Cardinal Health Inc.      & 0.01  & 0.03&    0.01  &   0.07  &&Grainger Inc.&   0.00 & 0.03&  0.00  &  0.00  \\          
Caterpillar Inc.          &  0.01 & 0.06&   0.08   &  0.13   &&Halliburton Co.& 0.00  &0.00& 0.00 & 0.00 \\          
Chubb Corp.               &  0.00  &0.00 &  0.00  &  0.00    &&Hasbro Inc.&   0.05   & 0.08 &0.28&0.18\\          
Coca-Cola Enterprises     &  0.00  &0.00 &  0.00  &  0.00    &&Health Care REIT& 0.00  &0.00& 0.00 & 0.00 \\          
Carnival Corp.            &  0.00  &0.00 &  0.00  &  0.00    &&Home Depot&   0.00    & 0.07  & 0.00 & 0.01\\       
CIGNA Corp.               &  0.04 &0.00&    0.66  &  0.93    &&Hess Corporation&  0.83    & 0.33 &0.92&0.97\\
Cincinnati Financial      &  0.00  &0.00 &  0.00  &  0.00    &&Harley-Davidson&  0.02    &  0.00 &0.58&0.92\\
Clorox Co.                &  0.00  &0.00 &  0.00  &  0.00    &&Hewlett-Packard&   0.00  & 0.04 &0.00&0.00\\
Comerica Inc.             &  0.00  &0.00 &  0.00  &  0.04    &&Block H and R& 0.00  & 0.01 &0.00&0.00\\
CMS Energy                &  0.00 &0.00&    0.00  &  0.00    &&Hormel Foods Corp.&   0.01   & 0.06 &0.00&0.04\\
CenterPoint Energy        &  0.00 &0.00&    0.00  &  0.00    &&The Hershey Company&  0.13    & 0.27 &0.56&0.20\\
Cabot Oil and Gas         &  0.00 &0.01&    0.00  &  0.00    &&Intel Corp.&    0.00   & 0.10 &0.00&0.00\\
ConocoPhillips            &  0.23 & 0.19&   0.42   &  0.44   &&International Paper&    0.00       & 0.29 &0.00&0.00\\
Campbell Soup             &  0.00 &0.00&    0.00  &  0.00    &&Interpublic Group&0.00  &0.03& 0.00 & 0.00 \\
CSX Corp.                 &  0.02 &0.15&   0.03   &  0.03    &&Ingersoll-Rand PLC&0.02  &0.10& 0.50 & 0.44\\
Cablevision Corp.         &  0.00 &0.00&    0.00  &  0.00    &&Johnson Controls&  0.00    & 0.01 &0.33&0.01\\
Chevron Corp.             & 0.00   & 0.03&   0.01   &  0.01  &&Jacobs Eng. Group&   0.22   & 0.07 &0.35&0.68\\
Dominion Resources        &  0.00 &0.00&    0.00  &  0.00    &&Johnson and Johnson&0.00 &0.00&    0.00  &  0.00 \\

\bottomrule
\end{tabular}
\end{table}

From the results in {Table} \ref{tab:table4}, we found that the linearity assumption is rejected by the proposed tests in $81 (88.0\%)$ cases compared with $64 (69.6\%), 74 (80.4\%)$, and $71 (77.2\%)$ cases on the basis of the test statistics $Q^\star,Q_{22}$, and $Q_{\textrm{WL}}$, respectively.  
This arguably suggests that the proposed tests are preferable to test the presence of nonlinearity in AR models for asset returns.

%%==============================================================================================================%%
\subsection{\label{sec:application2}Goodness-of-fit-tests for nonlinear time series models}
%%==============================================================================================================%%

\begin{figure}[!htbp]
\centering
\includegraphics[width=0.9\textwidth,height=0.25\textheight]{%figures/
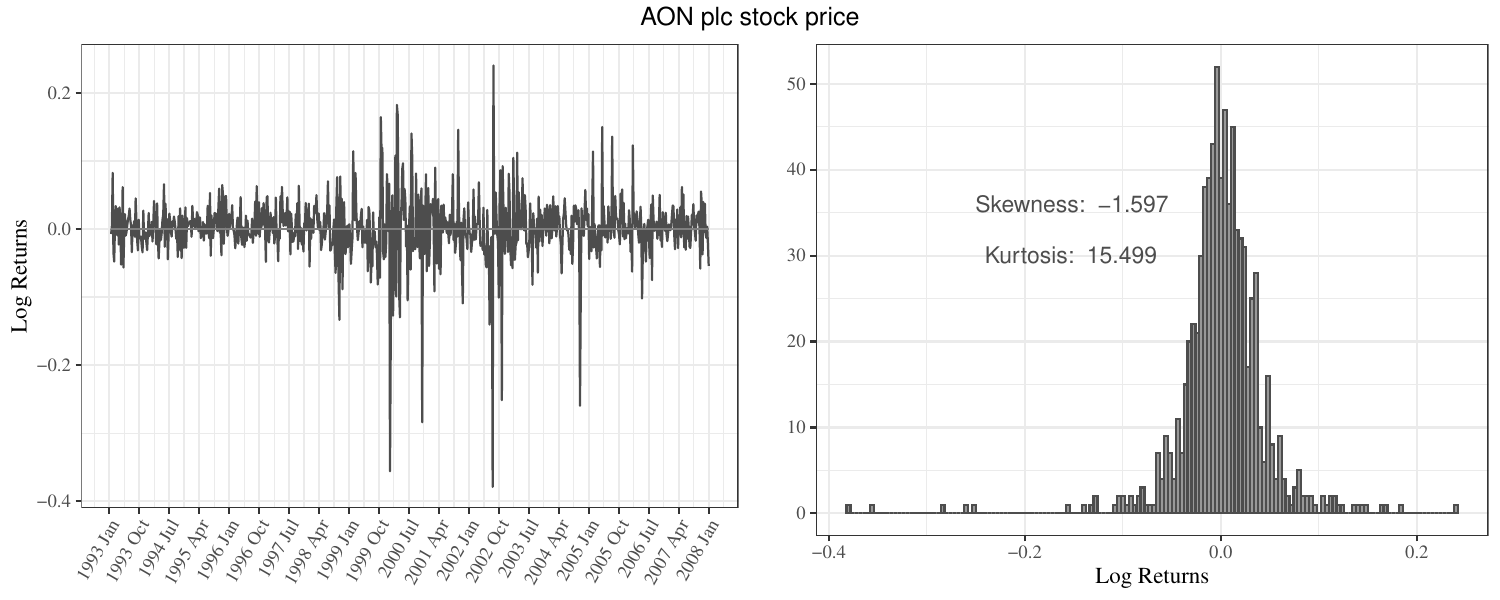}
\caption{Time series of weekly closing returns of the AON plc company from 18  June 1993 to 31 December 2007 (left) and histogram of their log-returns.}
\label{fig:AON}
\end{figure}

We examine the ability of the portmanteau tests to distinguish an unsuitable model for weekly stock returns of Aon plc company studied in Section \ref{sec:application1}. 
We find a strong evidence against linearity in the AR model for the returns of this company by using portmanteau tests. 
The Aon plc returns are displayed in Figure \ref{fig:AON} which shows that the 
log-return series have high persistence in volatility with negative skewness and excess kurtosis.
We conclude, therefore, that these returns might exhibit conditional heteroskedasticity effects, and a model that belongs to the ARCH family with a Student-$t$ distribution of the error process might better explain the leptokurtic distribution of the returns.
Thus, we fit ARCH(1), AR(1)-ARCH(1), ARCH(2), and GARCH(1,1) models and apply the test statistics $C^\star, Q_{\textrm{WL}}$, and $Q_{\textrm{LM}}$ at lag value $m=6$. 
The asymptotic p-values for testing the model adequacy based on these proposed statistics are reported in Table \ref{tab:AON}.
From this table, the test statistic $Q_{\textrm{LM}}$  fails to detect the inadequacy in all of the fitted models, whereas the test statistic $Q_{\textrm{WL}}$ 
suggests that the ARCH(1), ARCH(2) and GARCH(1,1) models might be suitable to describe for the Aon plc returns.
Only the proposed test statistics suggest a clear indication of inadequacy of the ARCH(1), AR(1)-ARCH(1), and ARCH(2) models, while the GARCH(1,1) model might be an adequate model for the Aon plc returns according to the proposed test statistics. 

\begin{table}[htb!]
\centering
\small
\renewcommand{\arraystretch}{0.7}
\caption{\label{tab:AON} P-values for testing the adequacy of the indicated model for Aon plc returns based on the listed statistic.
}
\begin{tabular}{l llll}
\toprule
Fitted model&&$C^\star$& $Q_{\textrm{WL}}$& $\textrm{LM}$ \\
\midrule
ARCH(1) && 0.003&  0.049   &  0.137\\
AR(1)-ARCH(1) && 0.001&  0.025    &  0.126\\
ARCH(2) && 0.008&  0.076   &  0.125\\
GARCH(1,1) && 0.208&  0.467  &  0.658\\ 
\bottomrule
\end{tabular}
\end{table}

%%==============================================================================================================%%
\section{\label{sec:discussion}Conclusions}
%%==============================================================================================================%%

In this article, we have introduced four mixed portmanteau statistics  for assessing the adequacy of time series models. 
The tests we propose are based on a linear combination of three auto-and-cross-correlation components.  
The first and second components are derived from the autocorrelations of residuals and their squared values, respectively. Meanwhile, the third component takes into account the cross-correlations between the residuals and their square values, considering both positive and negative lags.
Two of these tests can be viewed as an extended version of the \citeauthor{LjungBox1978} test, while the others can be considered an extension of the \citeauthor{BoxPierce1970} test. Based on our simulation study, it is recommended to use the proposed tests, which can be seen as an extended version of the \citeauthor{LjungBox1978} test. These tests demonstrate better control over the type I error probability compared to existing tests. Furthermore, they generally exhibit more statistical power than tests relying on the statistics introduced by \cite{LiMak1994}, \cite{McLeodLi1983}, \cite{ZM2019}, and \cite{Wong2005}.

Simulation results indicate that combining $\bm {R}{(1,2)}$ and $\bm {R}{(2,1)}$ in a test statistic significantly reduces the test's power. 
This can be justified by the lack of independence between these two distinct components, as they share a substantial amount of information about correlation. 
Consequently, they will add complex redundant correlation which leads to decrease the power in the proposed test.

Some of the test statistics that we have discussed have high computational burdens, 
so we have implemented them in an {\tt R} package named {\tt portes} \citep{MahdiMcLeodportes2020}.
The idea discussed in this article may be extended to formulate an omnibus portmanteau test that combines the cross-correlations between the residuals and their square values at both positive and negative lags with the autocorrelations of the residuals and their squared values.
The framework we propose could be expanded to identify seasonality in time series and to detect various types of nonlinearity dependence in multivariate time series, as discussed by \citep{Mahdi2016}.

In this article, our focus has been on measures derived from second-order mixed moments. Nevertheless, there is potential for extending these measures to higher-order moments. Our simulation study revealed that severe skewness, such as in the case of a skewed $t$-distribution, can distort the size of all portmanteau tests. When distributional assumptions are relaxed, the robustness of bootstrapping and Monte Carlo significance test approaches becomes evident \citep{EfronTibshirani1994, LinMcLeod2006, MahdiMcLeod2012}. Therefore, a possible extension of this article could involve considering these approaches for calculating p-values.

%%==============================================================================================================%%
\section{\label{sec:Appendix}Appendix}
%%==============================================================================================================%%

%%==============================================================================================================%%
%\subsection{\label{sec:autocross}Auto-and-cross-correlations of residual and their squares}
%%==============================================================================================================%%

%%==============================================================================================================%%
%\subsection{\label{sec:asymptot} Asymptotic distribution}
%%==============================================================================================================%%
By Taylor's theorem there exists a random vector $\tilde{\bm{\theta}}_n$ on the line segment between $\bm{\theta}_0$ and $\bm{\hat{\theta}}_n$ such that
\begin{equation}\label{Taylor00}
	\widehat{\bm \Gamma}_{(1,1)}=\bm \Gamma_{(1,1)} + \frac{\partial \bm \Gamma_{(1,1)}}{\partial\bm{\theta}^\top}(\widehat{\bm{\theta}}_n-\bm{\theta}_0) +\frac{1}{2} (\widehat{\bm{\theta}}_n-\bm{\theta}_0)^\top\frac{\partial^2 \bm \Gamma_{(1,1)}(\tilde{\bm{\theta}}_n)}{\partial\bm{\theta}^\top\partial\bm{\theta}} (\widehat{\bm{\theta}}_n-\bm{\theta}_0),
\end{equation}
where the second derivative $\dfrac{\partial^2 \bm \Gamma_{(1,1)}(\tilde{\bm{\theta}}_n)}{\partial\bm{\theta}^\top\partial\bm{\theta}}$ depends on the second-order derivatives $\dfrac{\partial^2\gamma_{(1,1)}(k)}{\partial\bm{\theta}^\top\partial\bm{\theta}}$. 
By assumption, the second-order partial derivatives are dominated by a fixed integrable function for every $\bm{\theta}$ in a ball $B$ around $\bm{\theta}_0$, so that the probability of the event $\left\{\widehat{\bm{\theta}}_n \in B\right\}$ tends to 1. 
Thus
\begin{equation}\label{Taylor0}
	\widehat{\bm \Gamma}_{(1,1)}=\bm \Gamma_{(1,1)} + \frac{\partial \bm \Gamma_{(1,1)}}{\partial\bm{\theta}^\top}(\widehat{\bm{\theta}}_n-\bm{\theta}_0) + \frac{1}{2} (\widehat{\bm{\theta}}_n-\bm{\theta}_0)^\top O_p(1) (\widehat{\bm{\theta}}_n-\bm{\theta}_0).
\end{equation}	
As the sequence $\left(\widehat{\bm{\theta}}_n-\bm{\theta}_0\right) O_P(1)=o_P(1) O_P(1)$ converges to 0 in probability when $\widehat{\bm{\theta}}_n$ is consistent for $\bm{\theta}_0$, we can rewrite (\ref{Taylor0}) by employing the first-order Taylor series approximation:
\begin{equation}\label{Taylor11}
	\widehat{\bm \Gamma}_{(1,1)}\approx \bm \Gamma_{(1,1)} + \frac{\partial \bm \Gamma_{(1,1)}}{\partial\bm{\theta}^\top}(\widehat{\bm{\theta}}_n-\bm{\theta}_0),
\end{equation}
where
\begin{eqnarray*}
\frac{\partial \bm \Gamma_{(1,1)}}{\partial\bm{\theta}^\top}&=&\bigg(\frac{\partial \gamma_{(1,1)}(1)}{\partial\bm{\theta}^\top},\cdots,\frac{\partial \gamma_{(1,1)}(m)}{\partial\bm{\theta}^\top}\bigg)^\top,\\
\frac{\partial \gamma_{(1,1)}(k)}{\partial\bm{\theta}^\top}&=&-\frac{1}{n}\sum_{t=k+1}^{n}\frac{\varepsilon_{t}}{2h_{t}^{3/2}}\frac{\partial h_{t}}{\partial\bm{\theta}^\top}\frac{\varepsilon_{t-k}}{\sqrt{h_{t-k}}} -\frac{1}{n}\sum_{t=k+1}^{n}\frac{1}{\sqrt{h_{t}}}\frac{\partial \mu_{t}}{\partial\bm{\theta}^\top} \frac{\varepsilon_{t-k}}{\sqrt{h_{t-k}}}\\
&&-\frac{1}{n}\sum_{t=k+1}^{n}\frac{\varepsilon_{t-k}}{2h_{t-k}^{3/2}}\frac{\partial h_{t-k}}{\partial\bm{\theta}^\top}\frac{\varepsilon_{t}}{\sqrt{h_{t}}} -\frac{1}{n}\sum_{t=k+1}^{n}\frac{1}{\sqrt{h_{t-k}}}\frac{\partial \mu_{t-k}}{\partial\bm{\theta}^\top}\frac{\varepsilon_{t}}{\sqrt{h_{t}}}.
\end{eqnarray*}
By the ergodic theorem, for large $n$, and after taking the expectation with respect to $\mathcal{F}_{t-1}$, it is straightforward to show that
$$
\frac{\partial \gamma_{(1,1)}(k)}{\partial\bm{\theta}^\top}\overset{\mathrm{a.s.}}{\to}-\widetilde{\bm X}_{11}(k),
$$
where 
$$
\displaystyle{\widetilde{\bm X}_{11}(k)=\mathrm{E}\bigg(\frac{1}{\sqrt{h_{t}}}\frac{\partial \mu_{t}}{\partial\bm{\theta}^\top} \frac{\varepsilon_{t-k}}{\sqrt{h_{t-k}}}\bigg)}
$$
is a $1\times l$ vector, which can be consistently estimated by 
\begin{equation}
\bm X_{11}(k)=\frac{1}{n}\sum_{t=k+1}^{n}\frac{1}{\sqrt{\widehat{h}_{t}}}\frac{\partial \mu_{t}}{\partial\bm{\theta}^\top} \frac{\widehat{\varepsilon}_{t-k}}{\sqrt{\widehat{h}_{t-k}}}.
\end{equation}
It follows that $\widehat{\bm \Gamma}_{(1,1)}$ stated in \eqref{Taylor11} can be expressed as
$\widehat{\bm \Gamma}_{(1,1)}\approx \bm \Gamma_{(1,1)} - \bm X_{11}(\widehat{\bm{\theta}}_n-\bm{\theta}_0)$, 
where $\bm X_{11}=(\bm X_{11}^\top(1),\cdots, \bm X_{11}^\top(m))^\top$ is the resultant $m\times l$ matrix.
Thus, by scaling each term by the variance of the standardized residual, we have  
\begin{align}\label{eq:r11}
\sqrt{n}\widehat{\bm R}_{(1,1)} & \approx \sqrt{n}(\widehat{r}_{(1,1)}(1),\cdots,\widehat{r}_{(1,1)}(m))^\top\nonumber\\
& = \sqrt{n}(\rho_{(1,1)}(1),\cdots,\rho_{(1,1)}(m))^\top - \bm X_{11}\sqrt{n}(\widehat{\bm{\theta}}_n-\bm{\theta}_0).
\end{align}
When $\xi_t$ is normally distributed, the random vector $\sqrt{n}\widehat{\bm R}_{(1,1)}$ is asymptotically
normal distributed with a mean of zero vector and a variance $\bm I_{m}-\bm X_{11}\bm \Sigma^{-1}\bm X_{11}^\top$, where $\bm I_{m}$ is the identity $m\times m$ matrix.

For the case of $\widehat{\bm\Gamma}_{(2,2)}$ and $\widehat{\bm R}_{(2,2)}$, \cite{LiMak1994} and \cite{LingLi1997} showed that 
\begin{align}\label{eq:r22}
\sqrt{n}\widehat{\bm R}_{(2,2)}&\approx\sqrt{n}(\widehat{r}_{(2,2)}(1),\cdots,\widehat{r}_{(2,2)}(m))^\top \nonumber\\
&= \sqrt{n}(\rho_{(2,2)}(1),\cdots,\rho_{(2,2)}(m))^\top - \frac{1}{\sigma^2}\bm X_{22}\sqrt{n}(\widehat{\bm{\theta}}_n-\bm{\theta}_0),
\end{align}
where 
$\bm X_{22}=(\bm X_{22}^\top(1),\cdots, \bm X_{22}^\top(m))^\top$ is an $m\times l$ matrix, and $\bm X_{22}(k)$ is given by
\begin{equation}
\bm X_{22}(k)=\frac{1}{n}\sum_{t=k+1}^{n}\widehat{h}_{t}^{-1}\frac{\partial h_{t}}{\partial\bm{\theta}^\top}\left(\frac{\widehat{\varepsilon}_{t-k}^2}{\widehat{h}_{t-k}}-1\right).
\end{equation}
The authors proved that $\sqrt{n}\widehat{\bm R}_{(2,2)}$ is asymptotically
normal distributed with a mean of zero vector and a variance $\bm I_{m}-\frac{1}{4}\bm X_{22}\bm \Sigma^{-1}\bm X_{22}^\top$.

Now, we consider the case that $r=1$ and $s=2$.
Analogous to the reasoning in equations (\ref{Taylor00}-\ref{Taylor11}), we can express the first-order Taylor series approximation of $\widehat{\bm \Gamma}_{(1,2)}$ as follows:
\begin{equation}\label{Taylor12}
\widehat{\bm \Gamma}_{(1,2)}\approx \bm \Gamma_{(1,2)} + \frac{\partial \bm \Gamma_{(1,2)}}{\partial\bm{\theta}^\top}(\widehat{\bm{\theta}}_n-\bm{\theta}_0),
\end{equation} 
where 
\begin{eqnarray*}
\frac{\partial \bm \Gamma_{(1,2)}}{\partial\bm{\theta}^\top}&=&\bigg(\frac{\partial \gamma_{(1,2)}(1)}{\partial\bm{\theta}^\top},\cdots,\frac{\partial \gamma_{(1,2)}(m)}{\partial\bm{\theta}^\top}\bigg)^\top,\\
\frac{\partial \gamma_{(1,2)}(k)}{\partial\bm{\theta}^\top}&=&-\frac{1}{n}\sum_{t=k+1}^{n}\frac{\varepsilon_{t}}{2h_{t}^{3/2}}\frac{\partial h_{t}}{\partial\bm{\theta}^\top}\left(\frac{\varepsilon_{t-k}^2}{h_{t-k}}-1\right) -\frac{1}{n}\sum_{t=k+1}^{n}\frac{1}{\sqrt{h_{t}}}\frac{\partial \mu_{t}}{\partial\bm{\theta}^\top} \left(\frac{\varepsilon_{t-k}^2}{h_{t-k}}-1\right)\\
&&-\frac{1}{n}\sum_{t=k+1}^{n}\frac{\varepsilon_{t-k}^2}{h_{t-k}^{2}}\frac{\partial h_{t-k}}{\partial\bm{\theta}^\top}\frac{\varepsilon_{t}}{\sqrt{h_{t}}} -\frac{1}{n}\sum_{t=k+1}^{n}\frac{2\varepsilon_{t-k}}{h_{t-k}}\frac{\partial \mu_{t-k}}{\partial\bm{\theta}^\top}\frac{\varepsilon_{t}}{\sqrt{h_{t}}}.
\end{eqnarray*}
By the ergodic theorem, for large $n$, note that
$$
\frac{\partial \gamma_{(1,2)}(k)}{\partial\bm{\theta}^\top}\overset{\mathrm{a.s.}}{\to}-\widetilde{\bm X}_{12}(k),
$$
where 
$$
\displaystyle{\widetilde{\bm X}_{12}(k)=\mathrm{E}\left(\frac{1}{\sqrt{h_{t}}}\frac{\partial \mu_{t}}{\partial\bm{\theta}^\top} \left(\frac{\varepsilon_{t-k}^2}{h_{t-k}}-1\right)\right)}
$$ 
is a $1\times l$ vector, which can be consistently estimated by 
\begin{equation}
\bm X_{12}(k)=\frac{1}{n}\sum_{t=k+1}^{n}\frac{1}{\sqrt{\widehat{h}_{t}}}\frac{\partial \mu_{t}}{\partial\bm{\theta}^\top} \left(\frac{\widehat{\varepsilon}_{t-k}^2}{\widehat{h}_{t-k}}-1\right).
\end{equation}
Thus,
 $\widehat{\bm \Gamma}_{(1,2)}$ stated in \eqref{Taylor12} may be expressed as
$\widehat{\bm \Gamma}_{(1,2)}\approx \bm \Gamma_{(1,2)} - \bm X_{12}(\widehat{\bm{\theta}}_n-\bm{\theta}_0)$, and
\begin{align}\label{eq:r12}
\sqrt{n}\widehat{\bm R}_{(1,2)} &\approx\sqrt{n}(\widehat{r}_{(1,2)}(1),\cdots,\widehat{r}_{(1,2)}(m))^\top\nonumber\\
&= \sqrt{n}(\rho_{(1,2)}(1),\cdots,\rho_{(1,2)}(m))^\top - \frac{1}{\sigma}\bm X_{12}\sqrt{n}(\widehat{\bm{\theta}}_n-\bm{\theta}_0),
\end{align}
where $\bm X_{12}=(\bm X_{12}^\top(1),\cdots,\bm X_{12}^\top(m))^\top$.

Similarly, for the case $r=2$ and $s=1$, it is straightforward to show that  
\begin{align}\label{eq:r21}
\sqrt{n}\widehat{\bm R}_{(2,1)} &\approx\sqrt{n}(\widehat{r}_{(2,1)}(1),\cdots,\widehat{r}_{(2,1)}(m))^\top\nonumber \\
&= \sqrt{n}(\rho_{(2,1)}(1),\cdots,\rho_{(2,1)}(m))^\top - \frac{1}{\sigma}X_{21}\sqrt{n}(\widehat{\bm{\theta}}_n-\bm{\theta}_0),
\end{align}
where $\bm X_{21}=(\bm X_{21}^\top(1),\cdots,\bm X_{21}^\top(m))^\top$, and $\bm X_{21}(k)$ is a $1\times l$ vector given by
\begin{equation}
\bm X_{21}(k)=\frac{1}{n}\sum_{t=k+1}^{n}\frac{1}{\sqrt{\widehat{h}_{t-k}}}\frac{\partial \mu_{t-k}}{\partial\bm{\theta}^\top} \left(\frac{\widehat{\varepsilon}_{t}^2}{\widehat{h}_{t}}-1\right).
\end{equation}
The assumptions on $\xi_t$ imply that the random vectors $\sqrt{n}\widehat{\bm R}_{(1,2)}$ and $\sqrt{n}\widehat{\bm R}_{(2,1)}$ are asymptotically
normal distributed with mean zero and variance $\bm I_{m}-\frac{1}{2}\bm X_{12}\bm \Sigma^{-1}\bm X_{12}^\top$ and $\bm I_{m}-\frac{1}{2}\bm X_{21}\bm \Sigma^{-1}\bm X_{21}^\top$, respectively.

%%==============================================================================================================%%
%\subsection{\label{sec:Theorem}Proof of Theorem 3.1}
%%==============================================================================================================%%

By utilizing the results from (\ref{eq:r11}), (\ref{eq:r22}), (\ref{eq:r12}), and (\ref{eq:r21}) we can deduce the joint distribution of $\widehat{\bm R}_{(1,1)},\widehat{\bm R}_{(2,2)},\widehat{\bm R}_{(r,s)}$, for the cases $r=1, s = 2$ and $r=2, s=1$.
Without loss of generality, we establish the proof for the case $r = 1$ and $s = 2$. 
The proof for $r = 2$ and $s = 1$ readily ensues.
Therefore, if the model defined in (\ref{AR.ARCH.model}) is correctly specified, we obtain:
$$
\sqrt{n}
\left( 
\begin{array}{c}
\widehat{\bm R}_{(1,1)} \\
\widehat{\bm R}_{(2,2)}\\
\widehat{\bm R}_{(1,2)}
\end{array}
\right)\approx\sqrt{n}\bm D
\left( 
\begin{array}{c}
\bm R_{(1,1)} \\
\bm R_{(2,2)}\\
\bm R_{(1,2)}\\
\displaystyle{\frac{1}{n}\frac{\partial\ell}{\partial\bm{\theta}}}
\end{array}
\right),
$$
where 
$$
\bm D=
\left( 
\begin{array}{cccc}
\bm I_{m} & \bm 0 & \bm 0 & -\bm X_{11}\bm \Sigma^{-1} \\
\bm 0 & \bm I_{m} & \bm 0 & -\frac{1}{\sigma^2}\bm X_{22}\bm \Sigma^{-1} \\
\bm 0 & \bm 0 & \bm I_{m} & -\frac{1}{\sigma}\bm X_{12}\bm \Sigma^{-1}
\end{array}
\right).
$$Note that $\{\varepsilon_t\}\overset{\mathrm{i.i.d.}}{\sim}\mathcal{N}(0,1)$ so that the factors $1/\sigma^2$ and $1/\sigma$ can be replaced by $1/2$ and $1/\sqrt{2}$, respectively.

Let ${\bm W_n=\sqrt{n}(\bm R_{(1,1)}^\top,  \bm R_{(2,2)}^\top, \bm R_{(1,2)}^\top,n^{-1}{\partial\ell}/{\partial\bm{\theta}^\top})^\top}$.
By using a martingale difference approach in terms of $\mathcal{F}_t$ and following the same arguments provided by \citet[Theorem 1]{Wong2005}, one can easily show that $\bm W_n\overset{\rm D}{\to}\mathcal{N}({\bm 0},{\bm V})$; 
hence, $\sqrt{n}({\widehat{\bm R}_{(1,1)}}^\top,{\widehat{\bm R}_{(2,2)}}^\top,{\widehat{\bm R}_{(1,2)}}^\top)^\top \overset{\rm D}{\to}\mathcal{N}_{3m}(\bm 0,\bm \Omega_{12})$,
where $\bm \Omega_{12}=\bm D\bm V\bm D^\top$.
The matrices $\bm D,\bm V$, and $\bm \Omega_{12}$ can be consistently estimated by their sample values, denoted by $\widehat{\bm D},\widehat{\bm V}$, and $\widehat{\bm \Omega}_{12}$, respectively.
Under the assumptions of the model stated in (\ref{AR.ARCH.model}), we get
\begin{equation*}
\widehat{\bm V}=
\left( 
\begin{array}{cccc}
\bm I_{m} & \bm 0 & \bm 0& \bm X_{11} \\
\bm 0 & \bm I_{m} & \bm 0&  \bm X_{22} \\
\bm 0 &\bm  0& \bm I_{m} &  \bm X_{12} \\
\bm X_{11}^\top & \bm X_{22}^\top & \bm X_{12}^\top&\bm \Sigma^{-1}
\end{array}
\right).
\end{equation*}
Thus, we reach
\begin{eqnarray*}
\widehat{\bm \Omega}_{12}&=&\widehat{\bm D}\widehat{\bm V}\widehat{\bm D}^\top\\
&\approx&
\left( 
\begin{array}{ccccc}
\bm I_{m}-\bm X_{11}\bm \Sigma^{-1}\bm X_{11}^\top& -(1/2) \bm X_{11}\bm \Sigma^{-1}\bm X_{22}^\top&  -(1/\sqrt{2}) \bm X_{11}\bm \Sigma^{-1}\bm X_{12}^\top \\
-(1/2)\bm  X_{22}\bm \Sigma^{-1}\bm X_{11}^\top & \bm I_{m}-({1}/{4})\bm X_{22}\bm \Sigma^{-1}\bm X_{22}^\top&   -(1/\sqrt{8}) \bm X_{22}\bm \Sigma^{-1}\bm X_{12}^\top \\
-(1/\sqrt{2}) \bm X_{12}\Sigma^{-1}\bm X_{11}^\top &   -(1/\sqrt{8}) \bm X_{12}\bm \Sigma^{-1}\bm X_{22}^\top & \bm I_{m}-({1}/{2})\bm X_{12}\bm \Sigma^{-1}\bm X_{12}^\top.
\end{array}
\right).
\end{eqnarray*}
For the ARMA models, we have $\bm X_{11}\approx \bm 0, \bm X_{22}=\bm 0$, and $\bm X_{12}=\bm 0$, whereas for GARCH models, we have $\bm X_{11}\approx 0$ and $\bm X_{12}=0$.
Furthermore, for large $n$, when the model stated in (\ref{AR.ARCH.model}) is correctly specified, the off-diagonal block matrices in the matrix $\widehat{\bm \Omega}_{12}$ are approximately zero.
Therefore, in general, the matrix $\widehat{\bm \Omega}_{rs}$ has the form stated as
\begin{equation}\label{gamma.hat1}
\widehat{\bm \Omega}_{rs}=
\left( 
\begin{array}{ccc}
\bm I_{m}-X_{11}\bm \Sigma^{-1}\bm X_{11}^\top& \bm 0& \bm 0\\
\bm 0& \bm I_{m}-\frac{1}{4}\bm X_{22}\bm \Sigma^{-1}\bm X_{22}^\top&  \bm 0\\
\bm 0&   \bm 0&\bm  I_{m}-\frac{1}{2}\bm X_{rs}\bm \Sigma^{-1}\bm X_{rs}^\top\\
\end{array}
\right),
\end{equation}
with $r\ne s\in \{1,2\}$.
\qedhere

\subsection*{Acknowledgment}

Our sincere thanks go to Dr. Ajay Jasra, the editor, Dr. Mathieu Gerber, the associate editor, the two anonymous reviewers, Dr. Jan G. De Gooijer, and Kazem Ghanbari for their valuable and insightful suggestions on the manuscript.

\subsection*{Disclosure statement}
The author reports no conflict of interest regarding this paper.

\bibliographystyle{chicago} 
\bibliography{mybibfile}%

\end{document}